\documentclass[fleqn]{2017SCGE}

\usepackage{graphicx,times}
\usepackage{natbib}
\usepackage{amssymb,amsmath}
\usepackage{aas_macros}
\usepackage{float}
\usepackage{mathrsfs}	
\usepackage{color}
\usepackage{multirow}
\usepackage{blindtext}
\usepackage{comment}
\usepackage{longtable,booktabs,supertabular}

\usepackage{setspace}

\def\gsim{\mathrel{\rlap{\raise0.7pt\hbox{$>$}}{\lower 0.8ex\hbox{$\sim$}}}}
\def\lsim{\mathrel{\rlap{\lower3.0pt\hbox{$\sim$}}\raise0.7pt\hbox{$<$}}}
            
\usepackage{color}

\usepackage{comment}

\begin{document}
\ensubject{subject}


\ArticleType{Article}
\SpecialTopic{SPECIAL TOPIC: }
\Year{2023}
\Month{04}
\Vol{xx}
\No{x}
\DOI{xxxx}
\ArtNo{000000}
\ReceiveDate{xxx}
\AcceptDate{xxx}
\title{Toward a stellar population catalog in the Kilo Degree Survey: the impact of stellar recipes on stellar masses and star formation rates}

\author[1,2]{Linghua Xie}{}%
\author[1,2]{Nicola R. Napolitano}{napolitano@mail.sysu.edu.cn}
\author[3]{Xiaotong Guo}{guoxiaotong@smail.nju.edu.cn}
\author[4]{Crescenzo Tortora}{}%
\author[5]{Haicheng Feng}{}
\author[1]{\\Antonios Katsianis}{}
\author[6,7]{Rui Li }{}
\author[1,2]{Sirui Wu}{}
\author[8]{Mario Radovich}{}
\author[9]{Leslie K. Hunt}{}
\author[2,10]{Yang Wang}{}
\author[2,11]{\\Lin Tang}{}
\author[1]{Baitian Tang}{}
\author[1,2]{Zhiqi Huang }{}


\AuthorMark{Xie L.}%
\AuthorCitation{Xie L. et al.}
\address[1]{School of Physics and Astronomy, Sun Yat-sen University, Zhuhai Campus, 2 Daxue Road, Xiangzhou District, Zhuhai, P. R. China;}
\address[2]{CSST Science Center for Guangdong-Hong Kong-Macau Great Bay Area, Zhuhai, China, 519082}
\address[3]{Institute of Astronomy and Astrophysics, Anqing Normal University, Anqing, Anhui 246133, China}
\address[4]{INAF -- Osservatorio Astronomico di Capodimonte, Salita Moiariello 16, 80131 - Napoli, Italy;}
\address[5]{Yunnan Observatories, Chinese Academy of Sciences, Kunming, 650011, Yunnan, People's Republic of China}
\address[6]{School of Astronomy and Space Science, University of Chinese Academy of Sciences, Beijing 100049, China;} 
\address[7]{National Astronomical Observatories, Chinese Academy of Sciences, 20A Datun Road, Chaoyang District, Beijing 100012, China}
\address[8]{INAF - Osservatorio Astronomico di Padova, via dell'Osservatorio 5, 35122 Padova, Italy}
\address[9]{INAF - Osservatorio Astronomico di Arcetri, Largo Enrico Fermi 5, 50125, Firenze, Italy}
\address[10]{Peng Cheng Laboratory, No.2, Xingke 1st Street, Shenzhen, 518000, P. R. China}
\address[11]{School of Physics and Astronomy, China West Normal University, ShiDa Road 1, 637002, Nanchong, China}

\abstract{The Kilo Degree Survey (KiDS) is currently the only sky survey providing optical ($ugri$) plus near-infrared (NIR, $ZYHJK_S$) seeing matched photometry over an area larger than 1000 $\rm deg^2$. This is obtained by incorporating the NIR data from the VISTA Kilo Degree Infrared Galaxy (VIKING) survey, covering the same KiDS footprint. As such, the KiDS multi-wavelength photometry represents a unique dataset to test the ability of stellar population models to return robust photometric stellar mass ($M_*$) and star-formation rate (SFR) estimates. Here we use a spectroscopic sample of galaxies for which we possess $u g r i Z Y J H K_s$ ``gaussianized'' magnitudes from KiDS data release 4. We fit the spectral energy distribution from the 9-band photometry using: 1) three different popular libraries of stellar {population} templates, 2) {single burst, simple and delayed exponential} star-formation history models, and 3) a wide range of priors on age and metallicity. 
As template fitting codes we use two popular softwares: {LePhare} and {CIGALE}.
We investigate the variance of the stellar masses and the star-formation rates from the different combinations of templates, star formation recipes and codes to assess the stability of these estimates
and define some ``robust'' median quantities to be included 
in the upcoming KiDS data releases. 
As a science validation test, we derive the mass function, the star formation rate function, and the SFR-$M_*$ relation for a 
low-redshift ($z<0.5$) sample of galaxies, that 
result in excellent agreement with previous literature data. The final catalog, containing $\sim290\,000$ galaxies with redshift $0.01<z<0.9$, is made publicly available.}

\keywords{Characteristics and properties of
external galaxies -- Stellar content and populations; Origin, formation, evolution, age, and star formation;   Masses and mass distribution}

\PACS{98.62.Lv,98.62.Ai,98.62.Ck}

\maketitle

\onecolumn

\section{Introduction}
\label{sec:intro}
The spectral energy distribution (SED) of galaxies provides crucial information on the properties of their stellar populations at the different cosmic epochs. In particular, the stellar mass content and the star formation history of galaxies are of major importance to understand the mechanisms of their formation, including the impact of the environment on their properties \cite[e.g.,][]{conselice+14}. For instance, the study of the stellar mass function as a function of the redshift is a crucial probe of the stellar mass assembly of galaxies \cite[e.g.,][]{2017A&A...605A..70D}, and combined with the halo mass function of simulations, can be used as a cosmological probe, e.g. in abundance matching studies (e.g. \cite{2009ApJ...696..620C}, \cite{Moster2013MNRAS.428.3121M}, \cite{Behroozi2019MNRAS.488.3143B}). Similarly, the star formation rate function can measure the growth of the stellar content of galaxies across the cosmic time (e.g. \cite{Dave2017}). A relevant example of scaling relation is the 
star formation versus stellar mass, also known as the galaxy main sequence (\cite{Brinchmann2004_SFR, 2007ApJ...670..156D, 2007ApJ...670..173D, 2007A&A...468...33E, 2007ApJ...660L..43N, 2007ApJ...660L..47N}). This is crucial to understand the formation mechanisms of galaxies, in particular the relation between the star formation activity across time (\cite{2009ApJ...701..787P, 2013ApJ...763..129S, 2015A&A...575A..74S, 2018A&A...619A..27B}), and the gas consumption during galaxy formation (\cite{2010ApJ...718.1001B, 2010ApJ...714L.118D, 2012MNRAS.421...98D, 2014MNRAS.444.2071D, 2016MNRAS.455.2592R, Hunt+20,Tortora+22,2022MNRAS.515.3249W}).    

The measurement of the galaxy stellar masses and star formation rates mainly relies on details of stellar population analyses (\cite{2005MNRAS.362..799Maraston05}, \cite{2016MNRAS.463.3409Vazdekis16}), and their ability to constrain the stellar mass-to-light ratios (e.g. \cite{1994ApJS...95..107W}) and specific star formation history (e.g. \cite{Boquien2019A&A...622A.103B_Cigale}). This is a notoriously complex problem (\cite{2010ApJ...712..833Conroy_Gunn}), due to the existence of degeneracies among some of the parameters, in particular dust, age and metallicity (e.g. \cite{1972A&A....20..361F}, \cite{1980ApJ...236..430O}, \cite{1994ApJS...95..107W}, \cite{2000ApJ...541..126M}).
Furthermore, in order to convert the stellar population parameters into ``galaxy'' properties, one needs to account for the galaxy intrinsic luminosity, which carries other uncertainties, e.g. galaxy distances, or redshifts. This step is generally incorporated in the stellar population codes that can model the SED using the redshift as a free parameter (e.g. \cite{2006A&A...457..841I}, \cite{Sawicki2012}, \cite{Acquaviva2015}, \cite{Wang2023}) or as an input from spectra or photo-z codes (e.g. \cite{2009ApJ...700..221FAST}, \cite{Alsing2023}, \cite{Leistedt2023}). 

Despite these difficulties, 
spectroscopical data (\cite{Vulcani2013}, \cite{Ness2016}, \cite{2014ApJ...788...72Gallazzi14}, \cite{Carnall2023})   
or multi-band photometry (e.g., \cite{2011MNRAS.418.1587Taylor_gama}, \cite{Bilicki2021A&A...653A..82B}) have been routinely used to derive stellar masses, age, metallicity using simple stellar population (SSP, e.g. \cite{Baldry2018MNRAS_GAMADR3}) or more complex stellar population models with a {parametrized} star formation history (SFH, e.g. delayed exponential: \cite{Boquien2019A&A...622A.103B_Cigale}, log-normal: \cite{Abramson2016}, double power law: \cite{Behroozi2013}, $\Gamma$: \cite{Katsianis2021}) or non-parametric SFHs \cite{Leja2022,Whitler2023}.

Optical broadband photometry alone cannot break the dust-age–metallicity degeneracies (e.g. \cite{1997A&A...320...41K}), while
extending the wavelength range in the near-infrared (NIR) can provide additional constraints that can alleviate them 
(\cite{1984AJ.....89.1300B, 1996AJ....111.2238P}, \cite{Boquien2019A&A...622A.103B_Cigale}). 
The combination of optical and NIR photometry is also effective for photometric redshifts from SED fitting techniques, which are an important ingredient in stellar population analyses. These consist of finding a model galaxy spectrum, given by a linear combination of representative stellar or galaxy templates, which best fits the observed galaxy SED (\cite{Bolzonella2000A&A, 2000ApJ...536..571B, 2006A&A...457..841I,Brammer2008}). Here, the wide baseline can alleviate the degeneracy between various galaxy spectra as a function of galaxy redshifts (\cite{2009ApJ...690.1236I, 2019MNRAS.486.5104L}). 

In this paper, we want to test the outcomes of different stellar population codes, namely {LePhare} (\cite{2006A&A...457..841I}) and {CIGALE} {\cite{Boquien2019A&A...622A.103B_Cigale}}, and different stellar population templates and star formation histories, using a multi-band, seeing matched catalog of galaxies collected in the fourth data release (DR4) of the Kilo Degree Survey (KiDS, \cite{Kuijken+19_KiDS-DR4}, K+19 hereafter). The catalog includes sources
for which we possess 1) optical photometry in $ugri$ bands and NIR photometry in $ZYHJK_s$ bands from the VISTA Kilo Degree Infrared Galaxy (VIKING, \cite{Edge+14_VIKING-DR1}), 2) spectroscopic redshifts (spec-$z$s, hereafter) from different surveys, and 3) deep learning photometric redshifts. It collects about 290\,000 sources, a subsample of which 
has already been used in KiDS to calibrate photometric redshifts (e.g., \cite{Hildebrandt2017MNRAS.465.1454H,2022A&A...666A..85Gaznet}). The advantage of spectroscopic redshifts is that they 
alleviate the degeneracies between colors and redshifts, which further impact the accuracy of the stellar parameters. The addition of photometric redshifts will also allow us to assess the impact of their larger uncertainties on the same stellar parameters.
In fact, {\it the final goal of this work is to evaluate the variance of the stellar population quantities from different SED fitting recipes, popular stellar population templates, as well as the uncertainties on redshifts.} We will determine what are the most stable parameters and define robust quantities suitable for science applications. This is a first step to define a strategy to produce a robust stellar population catalog for the upcoming KiDS data release 5 (KiDS-DR5, Wright et al. 2023).
The main parameters we are interested in are {the stellar mass and the star formation rate}, but we will also provide the catalog of ages and metallicities of the galaxy stellar populations from a large set of priors. Since for this spectroscopic sample we also possess very accurate morphotometric redshifts from deep learning (i.e. GaZNet, \cite{2022A&A...666A..85Gaznet}), we can finally test the impact of 
redshifts derived from 
pure multi-band photometric catalogs combining optical and NIR, like the ones expected to be collected from future large sky surveys like Euclid mission (\cite{Laureijs+11_Euclid}), Vera Rubin Legacy Survey in Space and Time (VR/LSST; \cite{Izevic+19_LSST}), China Space Station Telescope (CSST; \cite{Zhan+18_csst}).   

There have been previous works including stellar population analyses of KiDS galaxy catalogs, either determining stellar mass only, for weak lensing studies (\cite{2019A&A...632A..34Wright_KV450}) or estimating galaxy properties, including photometric redshifts and stellar masses, for bright galaxies (i.e. $r<21$, \cite{Bilicki2021A&A...653A..82B}), or estimating structural parameters and stellar mass to select ultra-compact and massive galaxies (\cite{Tortora+18_UCMGs,Scognamiglio+20_UCMGs}) and for central dark matter studies (\cite{Tortora+18_KiDS_DMevol}).
However, none of these has investigated the impact on the stellar masses of the combination of fitting procedure and stellar templates. A similar analysis has been provided for the CANDELS survey (\cite{2015ApJ...801...97Santini_candels}), where they {used} optical plus NIR photometry and tested the impact on stellar masses of different stellar population codes, stellar templates and star formation histories.

As a science validation test, we will conclude our analysis by using stellar mass and star formation rate estimates to derive the stellar mass function, the star formation rate function, and the mass vs. star formation rate relation of the galaxies from the KiDS spectroscopic sample, using both spectroscopic and deep learning redshifts and compare them with literature data at redshift $z<1$.

The paper is organised as follow. In Sect. \ref{sec:data} we introduce the data and the set-up of the stellar population analysis; in Sect. \ref{sec:results} we present the stellar population inferences, assess their accuracy and precision using a series of statistical estimators, and define a robust definition of the stellar mass and star formation estimates; in Sect. \ref{sec:discussion} we discuss the dependence of the accuracy and scatter on galaxy properties and finally show the galaxy mass function{, the star formation rate function,} and the stellar mass-star formation rate relation as a science validation test; in Sect. \ref{sec:concl} we draw some conclusions and perspectives for future analyses.
Throughout the paper, we will adopt the following cosmological parameters: $\Omega_m = 0.3$, $\Omega_\Lambda = 0.7$, $H_0 = 70 \rm \, km  s^{-1} Mpc^{-1}$.

\section{Data and Methods}
\label{sec:data}
The spectroscopic sample which we use in this paper consists of 9-band photometry from the 1000 deg$^2$ {area} of KiDS data release 4 (KiDS-DR4 hereafter, see K+19), plus spectroscopic redshifts collected from the Galaxy Mass Assembly \cite[GAMA, ][]{Driver2011MNRAS_GAMA} survey, and the Sloan Digital Sky Survey/Baryon Oscillation Spectroscopic Survey \cite[SDSS/BOSS, ][]{boss2013}, overlapping with the KiDS footprint. We also add further machine learning redshifts from the GaZNet convolutional network presented in \cite{2022A&A...666A..85Gaznet}, as these have been demonstrated to provide very accurate redshifts up to $z\sim 3$, for galaxy samples with magnitude $r \lsim 22.5$. In the following we describe in more details the content of the dataset and the different stellar population model set-ups used to {\bf analyze} them.

\subsection{Photometry and spectroscopic redshifts}\label{sec:photo_spec}
The photometric data of the spectroscopic sample are collected from the KiDS and the VIKING surveys. These are two sister surveys covering a total area of  $1350$ deg$^2$ of the sky, in $ugri$ and $ZYJHK_s$ bands, respectively. 
The KiDS survey has been carried out at the VST/Omegacam telescope in Cerro Paranal (\cite{Capaccioli2011Msngr.146....2C}; \cite{Kuijken2011Msngr.146....8K}).
It has been optimized for weak lensing in the $r$-band, which provides best seeing imaging (average FWHM$\sim0.7''$), and mean limiting AB magnitude ($5\sigma$ in a $2''$ aperture) of $25.02\pm0.13$. The other bands have been observed with poorer seeing and reached mean limiting AB magnitudes of $24.23\pm0.12$, $25.12\pm0.14$, $23.68\pm0.27$ for $u, g$ and $i$, respectively (see K+19).

VIKING has been carried out at the VISTA/VIRCAM (\cite{Sutherland2015A&A...575A..25S_VISTA}) and 
complemented KiDS observations with five NIR bands ($Z, Y, J, H$ and $K$s). The median value of the seeing is $\sim 0.9''$ (\cite{Sutherland2015A&A...575A..25S_VISTA}), and the AB magnitude depths are 23.1, 22.3, 22.1, 21.5 and 21.2 in the five bands (\cite{Edge2013Msngr_VIKING}), respectively.

The 9-band fluxes have been measured via the Gaussian Aperture and PSF (GAaP) photometry method (\cite{2015MNRAS.454.3500K}), which gives colours that are corrected for PSF differences. Hence, GAaP photometry naturally provides seeing matched fluxes for each source in the catalog, by definition.
However, sources more extended than the aperture function result in underestimated total fluxes. In order to correct this systematic effect, a total aperture correction needs to be applied to derive the ``total'' galaxy properties (see Sect. \ref{sec:masses}).
As discussed in K+19 
the GAaP photometry is {Galactic} extinction corrected using
the \cite{SFD98_dust} maps with the \cite{SF11} coefficients.

As a spectroscopic database, we have collected redshifts from: 1) GAMA data release 4  (\cite{driver_2022_gama_dr4}), and 2) SDSS data release 17 (\cite{2022ApJS..259...35A}, SDSS-DR17 hereafter). Previous compilations of spectroscopic data overlapping with the KiDS area did not include SDSS-DR17, but included other high redshift datasets (see e.g. \cite{2022A&A...666A..85Gaznet} and reference therein). However, the statistics of galaxies matching the KiDS-DR4 catalog at redshift larger than $z\sim1$ is rather sparse. 
On the other hand, for the analysis we are interested to perform in this paper, SDSS-DR17 and GAMA provide a quite abundant sample of galaxies at $z\lsim1$.
In particular, GAMA is the most complete sample,
reaching $\sim95.5\%$ completeness for $r$-band magnitude $r<19.8$ (\cite{Baldry2018MNRAS_GAMADR3}).

To match the redshift distributions of the two catalogs, we exclude sources at $z>0.9$, where the overall catalog drops to a {constant} number of a few tens of galaxies per redshift bin, mainly from SDSS-DR17. We also notice that a large portion of sources at 
$z<0.005$ are classified as ``stars'' from their parent surveys. Hence, to avoid the contamination from other misclassified stars, we decide to 
use a conservative cut and select only sources with $z>0.01$. Equally, we exclude all sources classified as Quasars (QSO), as their SED might be dominated by the {nuclei} emission, rather than the stellar population light. These criteria together produce a final catalog of 
242678 GAMA and 77859 SDSS-DR17 galaxies, which includes 31728 repeated sources. For these duplicates, we adopt the SDSS-DR17 redshifts, which have errors, finally ending up with a total of 288\,809  objects.
In the following, we consider these sources to  be ``galaxies'', although we might still expect some minor contamination from unclassified QSO (or active galactic {nuclei}, AGN).

\begin{figure*}
    \hspace{-0.7cm} 
    \includegraphics[width=0.5\linewidth]
    {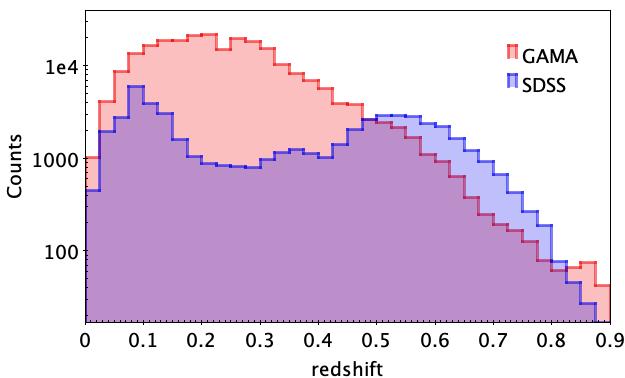}
    \includegraphics[width=0.5\linewidth]
    {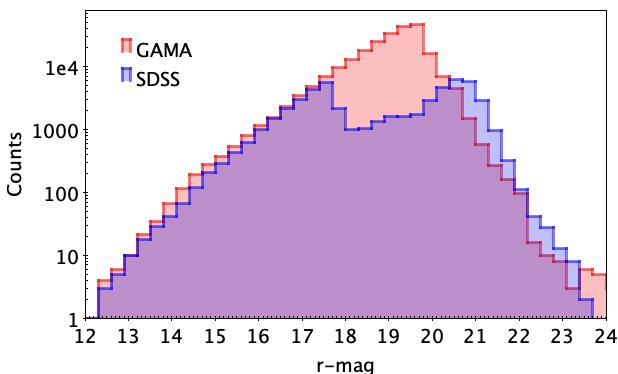}
    \caption{Properties of the spectroscopic sample. 
    Left: the distribution of the final selected ``galaxy'' sample (see text for details), separated in the different parent surveys. Right: the $r$-band MAG\_AUTO from the KiDS-DR4 catalog. 
    \label{fig:z_mag}
    }
\end{figure*}

The distributions of the redshift and the $r$-band Kron-like magnitude, MAG$\_$AUTO ($r$-mag for short), obtained by SExtractor \cite{Bertin_Arnouts96_SEx} for these galaxies are finally reported in Fig. \ref{fig:z_mag}, where we have broken the sample in the two original spectroscopic surveys, for clarity. From the $r$-mag distribution we can see the different completeness magnitude of the two samples, with SDSS-DR17 showing a peak at $r\sim17.8$. 
and GAMA 
at $r\sim19.8$.
The sample (in)completeness is not expected to impact the main goal of our analysis, which is to study the response of the 9-band optical+NIR photometry to the different stellar population recipes, however we will need to consider this when the stellar parameters will be used for the science validation test (see Sect. \ref{sec:sci_val}).

\subsection{Statistical estimators}
\label{sec:stat_est}
Here we introduce some statistical estimators we will use throughout the paper: 1) the relative bias, 2) the median {absolute} error and 3) the outlier fraction.\\ 
 \noindent
1)  The relative bias is defined as
\begin{equation}
     \Delta p = 
     p_{i}-r_{i},
     \label{eq:rel_bias}
\end{equation}  
where $p_{i}$ and $r_{i}$ are the estimated (log) parameters and the reference value for any $i$ galaxy of the sample. 
{In the case of redshifts, this becomes \begin{equation}
\mu=\frac{p_{i}-z_{i}}{1+z_{i}},
\end{equation} 
where $p_{i}$ are the predicted photometric redshifts and $z_{i}$ are the spectroscopic redshifts (see \cite{Amaro2021+photz}).}

\noindent
2) The Normalized median absolute deviation (NMAD) is then defined as: 
\begin{equation}
     {\rm NMAD} =  1.4826 \times {\rm median} ~(|BIAS - {\rm median} ~(BIAS)|).
\end{equation}
where we identify by BIAS either the $\Delta p$ or the $\mu$ defined above. This gives a measure of the overall scatter of the predicted values with respect to the 1-to-1 relation, i.e. the precision of the method.
~\\
3) Fraction of outliers.
{It is often useful to define the fraction of catastrophic estimates, that significantly deviate from the mean values, as a measure of the robustness of an estimator. In case of redshifts this is defined as the fraction of discrepant estimates,  
with the condition $|\mu|>0.15$ (see, e.g., \cite{2022A&A...666A..85Gaznet}). For the stellar population parameters we decided to use a 2$\sigma$ level in the log-normal distribution of the estimated values, which allow us to spot strong deviations from gaussianity.} 

\begin{figure*}
    \centering
    \includegraphics[width=0.6\linewidth]
    {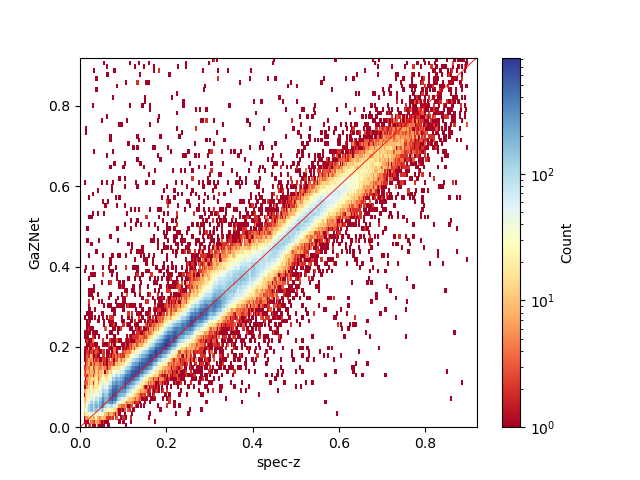}
    \caption{Morphoto-metric redshift from GaZNet vs. spectroscopic redshifts. {Data points are color-coded by the local density of data, according to the colorbar on the right. The diagonal red line shows the perfect 1-to-1 relation between the spec-$z$ on the X-axis and the morphoto-$z$ on the y-axis.}
    \label{fig:photo_spec_z}
    }
    \label{CNN_model_detail}
\end{figure*}

\subsection{Deep Learning morphoto-metric redshifts from GaZNet}
 As mentioned in Sect. \ref{sec:intro}, in this paper we want to test the robustness of the derived quantities from a full photometric samples. To do that, besides the spec-$z$ as in Sect. \ref{sec:photo_spec}, we use the morphoto-metric redshifts obtained by combining KiDS $r$-band images and the 9-band catalog using the Galaxy morphoto-Z Network (GaZNet, \cite{2022A&A...666A..85Gaznet}, Li+22 hereafter). 
 GaZNet has been previously tested on a KiDS galaxy sample (see Li+22 for details) and demonstrated to achieve very high precision in normalized median absolute deviation (NMAD=0.014 for $z\lsim1$ redshift and NMAD=0.041 for $z\gsim1$ redshift galaxies) and low outlier fraction ($0.4\%$ for lower and $1.27\%$ for higher redshift galaxies, respectively), down to $r\sim22$. These performances are better than the ones obtained by standard bayesian methods in KiDS for ``point'' estimates (e.g. BPZ, see \cite{2019A&A...632A..34Wright_KV450}) and other machine learning methods based on photometry only data applied previously to KiDS datasets (e.g. \cite{Cavuoti+15_KIDS_I}, \cite{Bilicki2021A&A...653A..82B}). 
 The level of accuracy reached by the deep learning estimates is shown in Fig. \ref{fig:photo_spec_z}, where we compare the GaZNet estimated {redshifts} vs. the spec-$z$ catalog described above. 
 In this figure we show the GaZNet estimates also for the SDSS-DR17 sample, that was not part of the deep learning training/testing in Li+22. As such, the SDSS-DR17 sample, added in this paper, represents a totally independent galaxy test sample with rather different distribution in redshift and luminosity than the original training sample (see Fig. \ref{fig:z_mag}). This gives us a more realistic sense of the scatter we can expect from the full photometric samples from KiDS, covering similar redshift/magnitude ranges.  For the predictions in Fig. \ref{fig:photo_spec_z}, we obtain a relative bias $\mu=0.005$, a NMAD=0.017 and an outlier fraction of 0.4\%, which are perfectly in line with the results found on \cite{2022A&A...666A..85Gaznet}, hence confirming the very good performances of the deep learning morphoto-$z$ provided by the GaZNet. We just notice a tail of outliers at $z\lsim0.05$, which are overestimated by the GaZNet and that might yet produce some systematics in the stellar population parameters.

\begin{table*}[t]
\footnotesize
\hspace{-1cm}
\begin{center}
\caption{Summary of the set-ups adopted for the stellar population models}
\begin{tabular}{lccccc}
\hline  
\hline\\
{\bf ID}  & \bf LP/BC03 &\bf LP/M05 &\bf LP/CB07 &\bf CI/BC03 &\bf CI/M05  \\
\\[0.01pt]
\hline
\\[0.1pt]
{\bf Code} &\multicolumn{3}{c}{LePhare} & \multicolumn{2}{c}{CIGALE} \\
\\[0.1pt]
{\bf Templates} & BC03 & M05 & CB07 & BC03 & M05 \\
\\[0.1pt]
{\bf IMF} & Chabrier & Kroupa$^{\ref{foot:IMF}}$ & Chabrier & Chabrier & Chabrier \\
\\[0.1pt]
{\bf SFH} & ExD & SB; ExD; SB+ExD & SB & DelEx & DelEx \\
\\[0.1pt]
{\bf $\tau$/Gyr} & \multicolumn{2}{c}{0.1, 0.3, 1, 2, 3, 5, 10, 15, 30} & $-$ &\multicolumn{2}{c}{0.25, 0.5, 0.75, 1, 1.5, 2, 2.5, 3, 4, 5} \\
\\[0.1pt]
{\bf Z [$10^{-2}$] }&  0.4, 0.8, 2
 & 0.4, 1, 2 (ExD) + 4 (SB)
 &  0.01, 0.04, 0.4, 0.8, 2, 5, {10} & 0.01, 0.04, 0.4, 0.8, 2, 5 &  0.1, 1, 2, 4  \\
\\[0.1pt]
{\bf(age/Gyr)} & \multicolumn{3}{c}{$0.5-13.5$} &  \multicolumn{2}{c}{0.5, 
 1, 1.5, 2, 2.5, 3, 4, 5, 6, 8, 10, 13} \\
step & \multicolumn{3}{c}{0.5} &  \\
\\[0.1pt]
{\bf Nebular Emission (NE)}& yes & yes & yes & yes & no \\
\\[0.1pt]
{\bf extinction law} & \multicolumn{5}{c}{Calzetti} \\
\\[0.1pt]
{\bf E(B-V)}& \multicolumn{3}{c}{0, 0.1, 0.2, 0.3} & \multicolumn{2}{c}{0.01, 0.02, 0.04, 0.06, 0.1, 0.2, 0.4, 0.6, 0.8} \\
\\[0.1pt]
  \hline
   \hline
   \\[0.1pt]
    \end{tabular}
    \label{tab:model_par}
    
\end{center}
\textsc{Note.} --- 
Row 1: ID prefix of the model: the full model is given by ID/SFH or ID/SFH/NE if nebular emissions (continuum and/or lines) are included. Row 2: code adopted as in Sects. \ref{sec:Lephare} and \ref{sec:cigale}. 
2: Initial Mass Function: note that we have converted the stellar masses from the Kroupa IMF to the Chabrier IMF by -0.05 dex (see footnote \ref{foot:IMF}). Row 3: different templates adopted: \cite[][BC03]{BC03}, \cite[][M05]{2005MNRAS.362..799Maraston05}, \cite[][CB07]{2007ASPC..374..303BC07}
Row 4: star formation history: SB=single burst, ExD=Exponential decline, DelEx=Delayed exponential.  Row 5: characteristic decaying time, as defined in Sects. \ref{sec:Lephare} and \ref{sec:cigale}. Row 6: metallicity values. Row 7: age range and steps. Note that the age cannot exceed the age of the universe at each redshift. Row 8: whether or not nebular lines are included in the model. Row 9: internal extinction law adopted. Row 10: the extinction values adopted in the models for the Calzetti law.  
\end{table*}

\subsection{{LePhare} stellar population: set-up and templates}
\label{sec:Lephare}
{LePhare} (\cite{1999MNRAS.310..540A, 2006A&A...457..841I}), is a template-fitting code, which performs a simple $\chi^2$ minimization between the stellar population synthesis (SPS) theoretical models and data, in a standard cosmology (see \S\ref{sec:intro}). 
In our analysis we adopt a \cite{Chabrier03} Initial Mass Function\footnote{In LePhare, there is no real option to set the IMF, but this is implemented in the stellar libraries. For the \cite{2005MNRAS.362..799Maraston05} libraries the IMF closer to Chabrier is the \cite{Kroupa01} IMF. To account for these IMF difference we will simply adopt the standard -0.05 dex correction to transform Kroupa-based into Chabrier-based masses. \label{foot:IMF}} (IMF), the \cite{Calzetti+94} dust-extinction law. We also include the contribution of nebular emission, e.g. from low-mass starforming galaxies (see Sect. \ref{sec:results}): LePhare uses a simple recipe based on the Kennicutt relations \cite{1998ApJ...498..541Kennicutt98} between the SFR and UV luminosity, H$\alpha$ and [OII] lines.
Regarding the stellar templates, we test three different libraries: 1) the standard \cite[][BC03 hereafter]{BC03}, 2) the \cite[][M05 hereafter]{2005MNRAS.362..799Maraston05} and 3) the \cite[][CB07 hereafter]{2007ASPC..374..303BC07} stellar population synthesis (SPS) models. We have also adopted three different models for the star formation history (SFH), $\psi(t)$: 1) a single burst (SB, hereafter), i.e. $\psi(t)=\delta(t_0)$, where $t_0$ is the age of the galaxy, 2) the exponentially declining law (ExD, hereafter), $\psi(t)\propto {\rm exp}(-t/\tau)$, and finally 3) a combination of both (SB+ExD), which is directly allowed by, e.g., the M05 stellar libraries.  
We remark here that the choice of the exponential declining SFH is due to the limited choice offered by Le Phare, even though the ExD is flexible enough to embrace a variety of realistic SFHs.
CIGALE (see below) will give us the chance to make a different choice, although a more general approach with a larger variety of SFHs will be considered in future analyses. 
The full LePhare set-up is summarized in Table \ref{tab:model_par}.
As anticipated in Sect. \ref{sec:intro}, we use the redshift, both spec-$z$ and morphoto-$z$, as input in {LePhare}. 
The stellar population parameters we use to perform the best fit to the GAaP 9-band magnitudes, described in Sect. \ref{sec:photo_spec}, are: age, metallicity, and star formation parameters (either $\delta(t_0)$ or $\tau$), which are assumed to vary as in Table \ref{tab:model_par}. Consistently with previous literature (e.g. \cite{2013A&A...556A..55I}, \cite{Bilicki2021A&A...653A..82B}), we use the best-fit parameters as a reference for this analysis.

\subsection{{CIGALE} stellar population: set-up and templates}
\label{sec:cigale}
We also adopt the Code Investigating GALaxy Emission (CIGALE, \cite{Boquien2019A&A...622A.103B_Cigale}, v2020.0), which can construct the FUV to the radio SEDs of galaxies and provide star
formation rate, attenuation, dust luminosity, stellar mass, and many other physical quantities, using composite stellar populations from simple stellar populations combined with highly flexible star formation histories.
For our analysis, we make use of BC03 and M05 stellar templates. Differently from {LePhare}, CIGALE does not have a pure ExD law among the SFH choices, hence we decide to adopt a delayed exponential law (DelEx, hereafter), $\psi(t)\propto t/\tau^2 {\rm exp}(-t/\tau)$, which is smoother than the exponential declining SFH from LePhare.
Consistently with {LePhare}, we have adopted a \cite{Chabrier03} Initial Mass Function (IMF), \cite{Calzetti+94} dust-extinction law and both the inclusion or not of nebular continuum and emission lines for the BC03 only. In CIGALE the nebular templates adopted are based on \cite{2011MNRAS.415.2920Inoue11}. The full set-up parameters, including the range of the stellar parameters adopted, {are} summarized in Table \ref{tab:model_par}. As for LePhare, we use the best-fit parameters from CIGALE in the following analysis.

\section{Results}
\label{sec:results}
In this section, we discuss the outcome of the different models summarized in Table \ref{tab:model_par}. These have, in some cases, very strong differences in the recipe of the star formation history (SFH), as we have adopted a single burst and both an exponentially declining and delayed exponential SFR, with a wide range of $\tau$ (see Sects. \ref{sec:Lephare} and \ref{sec:cigale}, and Table \ref{tab:model_par}).
This choice is made to explore the impact of  different SFHs on the stellar masses
and SFR estimates.
The SFH models above have have been effectively used to reproduce the properties of local galaxies \cite{Bell2000,Muzzin2013,Belli2019,Phillipps2019} and cosmic SFR density and stellar mass density at redshifts z $<$ 2 \cite{Nagamine2000,Mortlock2017,McLure2018}.
As anticipated, we also include the effect of  emission lines that, although they are generally important in massive galaxies at high redshift (e.g.\cite{2014A&A...563A..81deBarros_nebular}, \cite{2019A&A...631A.123Yuan_nebular}, but see also \cite{2015ApJ...803...11Stefanon_nebular}), can also be relevant for local low-mass starforming galaxies (e.g. \cite{2003A&A...401.1063Anders_EL_dwarfs}). 

Overall, 
the model combinations in Table \ref{tab:model_par} include a fair variety of libraries and SFHs, which we expect to provide realistic evidences of systematic effects. 
Moreover, as we are preparing the methods to be applied to the full KiDS photometric dataset, we will perform the same analysis using morphoto-$z$s as input, which will be provided to deeper limiting magnitudes that the ones offered by the spectroscopic ``galaxy'' sample (e.g. down to $r\sim22.5$ as seen in Li+22). This will allow us to evaluate the existence or not of systematics on stellar population parameters, and the impact on the precision of the estimates, due to the usage of the more scattered photometric redshifts.  
Once collected all the estimates from all the configurations in {Table 
 \ref{tab:model_par}}, we will 1) check the overall consistency among the different stellar parameters; 2) discuss the scatter of the parameters and possibly define some robust estimator for them. As mentioned in the Sect. \ref{sec:intro}, in this first paper we concentrate on the stellar masses and the star formation rates, as the most physical meaningful parameters one can extract from large multi-band photometric samples of galaxies, to study their evolution across the cosmic time. 
We use the estimates from BC03 templates and ExD star formation recipe in LePhare (LP/BC03/ExD in Table \ref{tab:model_par}) {\it as reference model for mass and star formation estimates}, if not otherwise specified. This is for uniformity with previous analyses in KiDS (e.g. \cite{2019A&A...632A..34Wright_KV450}).
To statistically assess the difference among the stellar mass and the SFR estimates among the different configurations, we will use the following estimators: 1) the relative bias, 2) the median {absolute} error and 3) the outlier fraction, defined in Sect. \ref{sec:stat_est}.

\begin{figure*}[t]
\centering 
\includegraphics[width=0.42\linewidth]{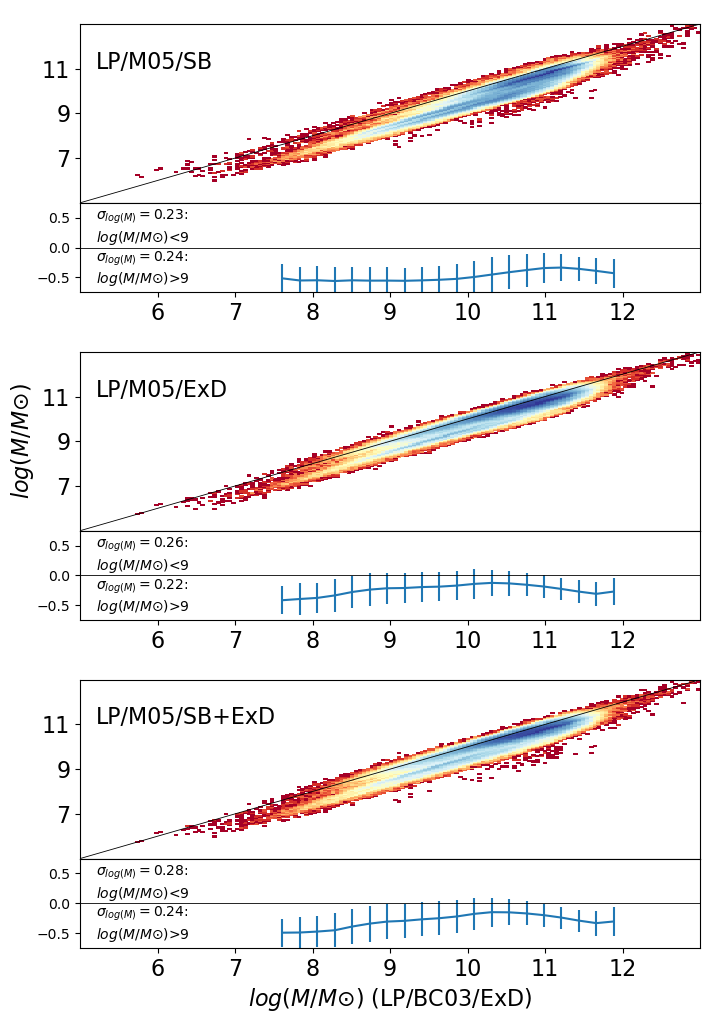}
\includegraphics[width=0.42\linewidth]{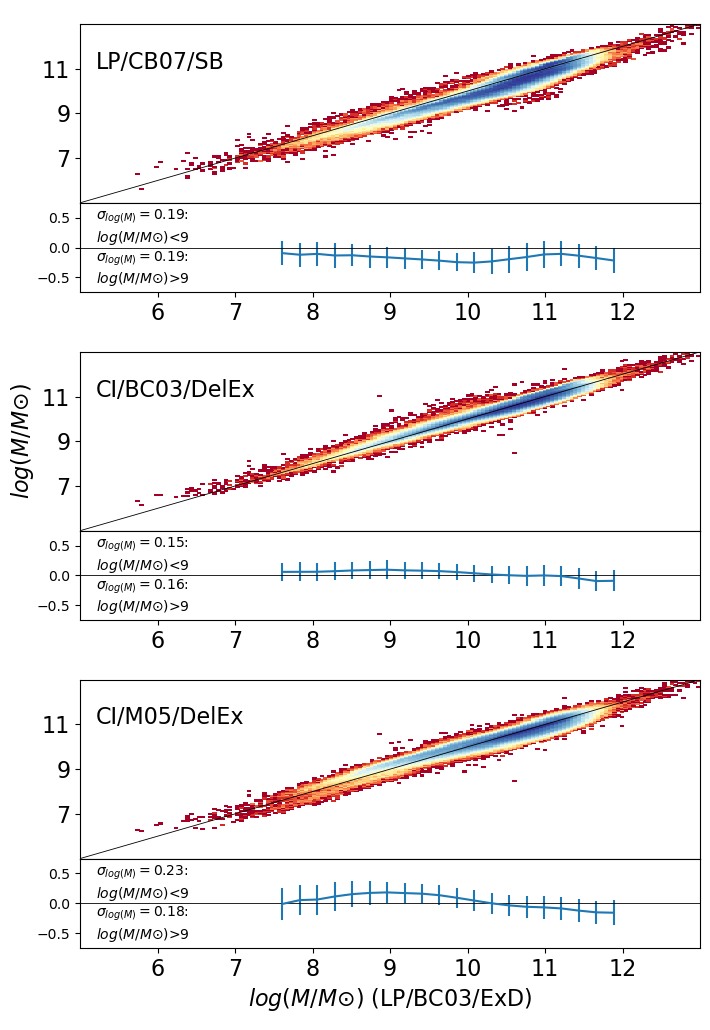}
    \caption{Stellar mass estimates from spectroscopic redshifts. Direct comparison between: 1) y-axis: the different configurations of code/stellar libraries/SFH as in Table \ref{tab:model_par} (see also legenda) and 2) x-axis: the reference configuration, LP/BC03/ExD, in the same Table. The bottom panels show the residuals, defined as $\Delta \log M=\log M_y-\log M_x$, with respect to the 1-to-1 relation, and the related scatter, defined as the standard deviation, $\sigma(\Delta\log M)$, in different mass bins.}
\label{fig:B03vsall}
\end{figure*}

\begin{table*}[t]
\footnotesize
\begin{center}
\caption{Stellar masses. Summary of the statistical estimator of the different configurations as in Table 
\ref{tab:model_par}.}
    \begin{tabular}{lcccccccc}
  \hline
   \hline
\\[0.1pt]
\bf Test& \bf LP/B03/ExD & \bf LP/M05/SB &\bf  LP/M05/ExD & \bf LP/M05/SB+ExD& \bf LP/CB07/SB & \bf CI/BC03/DelEx & \bf CI/M05/DelEx\\
  & \bf (NE) & \bf (NE) &\bf (NE)  & \bf \bf (NE) & \bf \bf (NE) & \bf \bf (NE) & \bf  \\
\\[0.1pt]
  \hline
\\[0.1pt]
\multicolumn{8}{c}{\bf Spectroscopic Redshifts}\\
\\[0.1pt]
\hline
\\[0.1pt]

Bias & -- & -0.423  & -0.178  & -0.207 &  -0.175 &  0.014 &  -0.009&\\
\\[0.1pt]
NMAD  & -- & 0.317 &  0.198 &  0.209 &  0.196 &  0.142  & 0.192 &\\
\\[0.1pt]
Out. frac. & -- & 3.9\% & 4.5\%  &   4.6\%   &   4.0\%   &   4.2\%  &    4.3\%
\\[0.1pt]
  \hline
\\[0.1pt]
\multicolumn{8}{c}{\bf Morphoto-metric Redshifts}\\
\\[0.1pt]
\hline
\\[0.1pt]
Bias & 0.033 &  -0.388 &  -0.142 &  -0.176 &  -0.135  & 0.050 &  0.033 &\\
 & (0.038)  & (-0.383)  & (-0.137)  & (-0.170) & (-0.130) & (0.056) &  -- \\
\\[0.1pt]
NMAD & 0.093  & 0.295  & 0.228  & 0.237 &  0.223 &  0.175 &  0.219 &\\
 & (0.078)  & (0.287)  & (0.229)  & (0.238) & (0.220) & (0.179) &  -- \\
\\[0.1pt]
Out. frac. & 3.8\% &  3.8\% 
 &  4.2\%  & 4.4\% &  3.6\%  &  3.2\% & 3.5\%   \\
 & (3.9\%) &  (3.8\%) &  (4.2\%) &  (4.4\%)  & (3.7\%) &  (3.2\%) &    -- \\
\\[0.1pt]

  \hline
   \hline
    \end{tabular}
    \label{tab:masses_stats}
    
\end{center}
\textsc{Note.} --- 
Statistical properties of the different model combinations from Table \ref{tab:model_par} as referred to the LP/BC03/ExD. {Results for the Nebular Emissions (NE) are given between brackets for the models using the GaZNets redshifts as input.}
\end{table*}

\subsection{Stellar masses}
\label{sec:masses}
In this section we show the results for the stellar masses for the case we fix the redshift of the galaxies of the sample to the spectroscopic and morphoto-metric redshifts, introduced in Sect. \ref{sec:photo_spec} and shown in Fig. \ref{fig:photo_spec_z}. By stellar masses, we aim at determining the total mass in stars, while we have seen in Sect. \ref{sec:photo_spec} that the seeing matched GAaP photometry adopted in KiDS does not correspond to a ``total aperture''. Hence, if using these fractional fluxes, the stellar masses calculated by the stellar population codes are 
the mass of stars required to produce the inputted
galaxy SED, resulting in {an} aperture bias.
Therefore, in order to recover a fair estimate of the total galaxy stellar mass, the observed SED must be representative of the total light emitted from the galaxy. 
In order to correct this systematic effect, we opt to use the quasi-total SExtractor, MAG\_AUTO, using the equation:

\begin{equation}
\rm M_{\rm *, corr}= M_{*,out}+0.4\times(GAAP\_r-
MAG\_AUTO)
\label{eq:corr}
\end{equation}

where $\rm M_{*,out}$ is the stellar mass estimated by the stellar population tools, GAAP\_r is the $r$-band GAaP magnitude from the KiDS catalog, and the $\rm M_{\rm *, corr}$ is the corrected ``total'' mass, under the assumption of constant mass-to-light ratios. In the following we will first show the results of the stellar population analysis using the spectroscopic redshift, then we compare these latters with the results of the morphoto-$z$ to estimate the impact of the larger uncertainties on these latter on determining galaxy distances (see Sect. \ref{sec:intro}). Finally, we discuss the impact of the inclusion of the nebular emissions in the models.

\subsubsection{Using Spectroscopic redshifts}
\label{sec:masses_spec}
We start showing the results obtained using the spectroscopic redshift as fixed parameter in the stellar population tools.
In Fig. \ref{fig:B03vsall}, we compare the stellar mass estimates from LePhare and CIGALE, using different libraries and SFHs and spectroscopic redshifts. All other parameters, in Table \ref{tab:model_par}, are kept varying in the model grid to be estimated via the SED fitting procedure. 
The range of masses is quite large and spans over almost 6 order of magnitudes from $\log M_*/M_\odot\sim6$ to $\log M_*/M_\odot\sim12$, although stellar masses $\log M_*/M_\odot\lsim7-7.5$ are compatible with globular cluster sized systems rather than galaxies. We cannot exclude the contamination from such compact stellar systems, but we decide to retain all sources in the catalog without making any mass based selection. Nonetheless, we will keep this cautionary note on the very low mass end in mind throughout the paper. 

Overall, the stellar masses all align along the 1-to-1 relation with residuals (bottom panels), defined as $\Delta \log M=\log M_y-\log M_x$, computed in different mass bins, that are generally distributed around zero, but with the LP models systematically smaller and the CI models rather aligned to the reference model, LP/BC03/ExD. All residuals, except LP/M05/SB, are consistent with zero within $1\sigma$ scatter, defined as the standard deviation of the $\Delta\log M$, $\sigma(\Delta\log M)$, at least for masses larger than $\log M/M_\odot\sim9$. In the same bottom panels, we report the mean scatter for the mass bins at $\log M/M_\odot<9$ and $>9$, showing generally a slightly larger values at lower masses (mean 0.22 dex) than larger masses (mean 0.20 dex), with the CI models also showing a systematically smaller $\sigma(\Delta\log M)$ than LP ones.

The bias, NMAD and outlier fraction of each configuration are summarized in Table \ref{tab:masses_stats}. 
Similarly to Fig. \ref{fig:B03vsall}, the bias is indeed consistent with zero for all configurations within the NMAD, except for LP/M05/SB for which the bias is statistically 
significant. CIGALE shows both a negligible bias and small NMAD, whether or not the same stellar libraries of the reference model from LePhare (BC03) are adopted, meaning that the code and the SFH can have an impact on the scatter but not on the accuracy of the stellar mass inferences. 
On the other hand, the large bias found for LP/M05/SB shows that the combination of template and SFH has a large impact on the bias, for a fixed fitting method. If we also fix the template (see e.g. M05), we can see that the bias can have rather large variations (from $-0.423$ of LP/M05/SB, to $-0.178$ of LP/M05/ExD), eventually due to the impact of the different SFH choices that exacerbate the difference on the treatment of thermally pulsating asymptotic giant branch (TP-AGB) phase by M05 (see e.g. \cite{Boquien2019A&A...622A.103B_Cigale}). Moreover, we notice a double sequence, at stellar masses $\log M_*/M_\odot\lsim10.8$ in the models including the exponential SFH, separating star-forming from quiescent galaxies. The same sequence is not evident on the SB model, which tends to assign younger ages and lower mass-to-light ratios to star-forming galaxies and ultimately ending into an overall strong underestimate of the stellar masses (see the negative biases in Table \ref{tab:masses_stats}). The CIGALE model using M05 and a delayed exponential (CI/M05/DelEx) shows a tighter distribution, with no sign of the double sequence. This confirms that the M05 models are more sensitive than others to the SFH, although there might be a residual component from the fitting (code) procedure, having CI models $\sim$30\% smaller scatter than the LP ones, on average. The NMAD generally mirrors these behaviors,
with M05 configurations being larger than the corresponding set-ups {from} other templates (see e.g. LP/M05/SB vs LP/CB07/SB or CI/M05/ExD vs CI/BC03/ExD).

All in all, {from} Fig. \ref{fig:B03vsall} we see that, using the spec-$z$ as input, the scatter of the different combinations are well confined within $\sim$0.2 dex and 
the outlier fraction is always very small ($\sim 4-5\%$), consistently with a log-normal distribution of the uncertainties with no pathological cases across the models.

\begin{figure*}[t]
\centering
    \hspace{0.cm} 
    \includegraphics[width=0.86\linewidth]{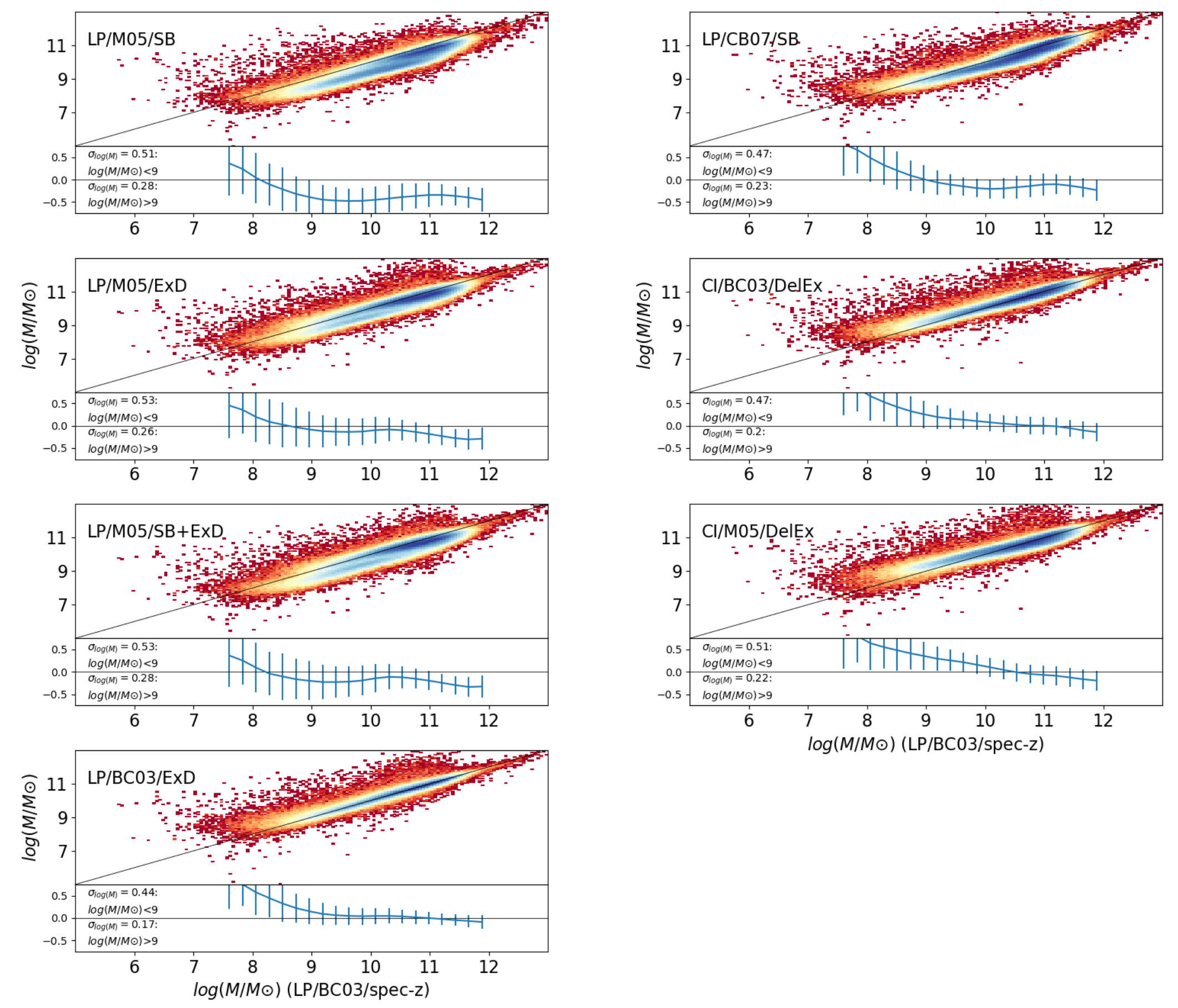}
\caption{Stellar mass estimates as for Fig. \ref{fig:B03vsall} but using the GaZNet morphoto-metric redshifts. }
    \label{fig:B03vsall_GAZ}
\end{figure*}

Considering the whole statistical estimators, we can conclude that stellar masses from spec-$z$ are a rather robust quantity with no signs of significant systematics, except for the LP/M05/SB model. 
This is consistent with findings from previous analyses also using optical + NIR photometry (e.g. Lee et al. 2010, \cite{2015ApJ...801...97Santini_candels}), although there are analyses reaching different conclusions (\cite{Maraston+10,Pforr+12}). 

\subsubsection{Using morphoto-metric redshifts}
\label{sec:mass_photoz}
We now show the results obtained using the GaZNet redshifts as fixed input in the stellar population tools. This is a critical test to check the impact of the use of noisier redshifts on the statistical estimators discussed in Sect. \ref{sec:stat_est}, and the overall variation of accuracy and precision of the estimates we might expect when applying this analysis to pure photometric datasets as the full KiDS photometric galaxy sample (see K+19 and future releases).

In Fig. \ref{fig:B03vsall_GAZ} we show the same correlations as in Fig. \ref{fig:B03vsall}, but using the GaZNet redshifts, while in Table \ref{tab:masses_stats} we report the corresponding statistical estimators. In this case, we also use the LP/B03/ExD model from the spec-$z$ as reference to check the impact of the GaZNet redshift in terms of accuracy and scatter. Basically, the results show that, for the same correlations seen in Fig. \ref{fig:B03vsall}, the relative bias of the different configurations is not worsened, meaning that the accuracy of the mass estimates is not affected by the use of the morphoto-$z$. This is eventually a consequence of the good accuracy of these latter as seen in Fig. \ref{fig:photo_spec_z}. On the other hand, we register an evident increase of the NMAD 
as a consequence of the morphoto-$z$ intrinsic statistical errors and outlier fractions, 
which is also mirrored by the scatter of the residual, at the bottom of the 1-to-1 relations, which is now of the order of 0.23 dex, for $\log M_*/M_\odot>9$, and 0.49 dex for $\log M_*/M_\odot<9$, on average. 
These large scatter at low stellar masses are mainly caused by 
the trend we see that below $\log M_*/M_\odot=8.5$, where stellar masses are systematically overestimated compared to those obtained with the spec-$z$. This is not an effect that comes from the particular set-up of the fitting procedure, as shown by the {comparison} of the LP/B03/ExD/morphoto-$z$ against the same set-up with spec-$z$
(bottom/left plot {in} Fig. \ref{fig:B03vsall_GAZ}). Even in this latter case, we see that below $\log M_*/M_\odot=8.5$ the positive bias is similar to the ones of all other configurations. We track the motivation of this systematics to some bias of the GaZNet redshifts for a group of objects at very low redshifts ($z<0.05$ see Fig. \ref{fig:photo_spec_z}), which turn-out to have also low masses. This can be due to some residual contamination from stars, not picked in the spectra classification, or just a failure of the GaZNet predictions at very low-$z$, which clearly impact the mass predictions. We will come back to this on Sect. \ref{sec:discussion}.
However, still looking at the LP/B03/ExD/morphoto-$z$ vs. spec-$z$, above 
$\log M_*/M_\odot=8.5$, the bias is almost absent and the only relevant effect is the GaZNet redshift scatter that, from the NMAD, is quantified in 0.09. 
This is confirmed by noticing that the general increase of the NMAD from the spectroscopic sample to the morphoto-metric sample, in Table \ref{tab:masses_stats}, is compatible with the sum in quadrature of the NMAD of the former with $0.09$ coming from the latter, consistently with some pseudo-Gaussian distributions. This is consistent with a log-normal distribution of the uncertainties of the stellar masses, which are confirmed by the outlier fractions that are all of the order of 5-6\% above 2$\sigma$ of the $\log M_*$ scatter. A more detailed discussion of the variation of the statistical estimators as a function of the sample properties is presented in Sect. \ref{sec:estim_galprop}.

\subsubsection{The impact of the nebular emissions on stellar masses}
\label{sec:NE_masses}
As anticipated at the beginning of Sect. \ref{sec:results}, we intend to check the impact of the inclusion of nebular emission on our models. Generally speaking, starforming galaxies can have their spectra heavily contaminated by nebular emissions. The most prominent ones are Ly$\alpha$ @$\lambda$1216\AA\, [OII] @$\lambda$3727\AA, H$\beta$ @$\lambda$4861\AA, [OIII] @$\lambda\lambda$ 4959\AA\ and 5007\AA,
H$\alpha$ $\lambda$6563\AA. These emissions are all sparsely distributed in the optical and NIR wavelength at redshift $z<1$, but they are generally fainter {than} the continuum collected by the broad bands in this redshift range, except for strong starburst, low-mass galaxies. Here, we have the chance to estimate the impact of the presence of these emissions on the stellar masses, while we will discuss the impact on the star formation rate estimates in Sect. \ref{sec:NE_sfr}. We consider the options offered by the {LePhare} and CIGALE (see details in Sects. \ref{sec:Lephare} and \ref{sec:cigale}) to implement the NE in the models as in Table \ref{tab:model_par}. The results of the statistical estimators are reported in Table \ref{tab:masses_stats}, between brackets, for all models considered. Here, we do not find any significant variation of the indicators of all models, which lets us to conclude that the stellar masses are poorly sensitive to the inclusion of the NE, regardless the stellar template, the SFH and the code adopted. We will keep the record of these models in the catalog and we will consider the for the discussion on the variance of the models in the discussion (Sect. \ref{sec:discussion}).

\subsection{Star Formation Rates}
\label{sec:sfr}
   
\begin{figure*}[t]
 \centering
    \hspace{-0.cm} 
    \includegraphics[width=0.5\linewidth]{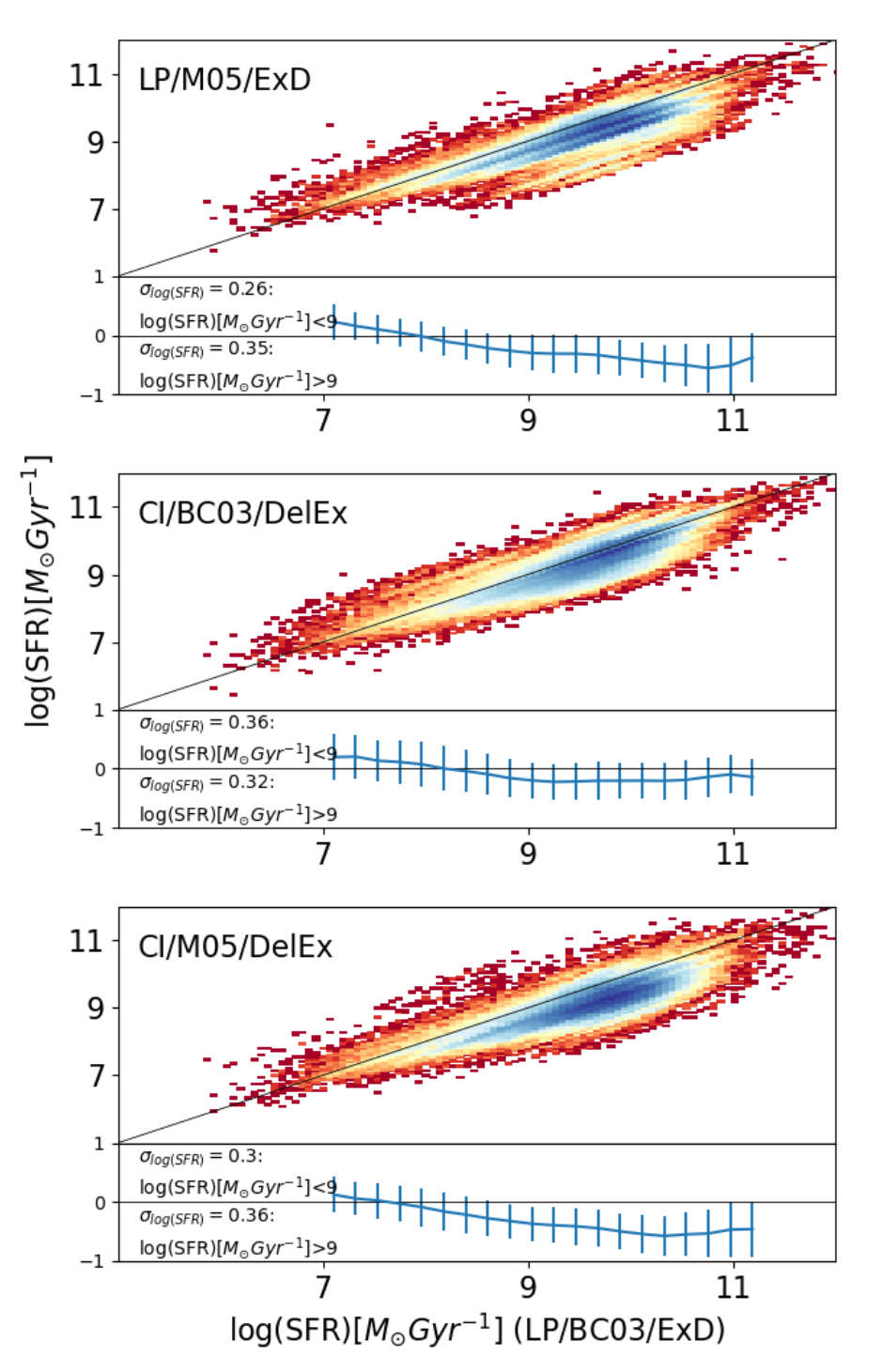}
    \caption{Star Formation Rates (SFR) from spectroscopic redshifts. Direct comparison between the different configurations of code/stellar libraries/SFH as in Table \ref{tab:model_par}  (see also the plots' legenda) and the reference configuration the LP/BC03/ExD in the same Table. The bottom panels show the residuals with respect to the 1-to-1 relation.
}
    \label{fig:B03vsall_SFR}
\end{figure*}
In this section, we present the results on the star formation rates. These measurements represent 
the current amount of stellar masses formed per unit of time corresponding to the best parameters of the assumed SFH model fitting the SED.
As, by definition, the single burst models do not provide any such estimate, they will be discarded in the following analysis. For the same reason, the mixed model allowed from the M05 libraries (SB+ExD) is almost equivalent to the ExD, as it returns the same SFR estimates for the galaxies best fitted with an exponential SFH (ExD). Hence, only the latter one will be listed in the result tables and figures for the LePhare models, together with the DelEx of CIGALE. 

We remind here the set of $\tau$ and ages adopted for the models in Table \ref{tab:model_par}. As seen, we have used a rather large sample of both parameters to check the impact of them on our inferences, even though some extreme values can be either slightly un-physical or too optimistic. For instance, there might be little sensitivity from the fitting procedure to effectively distinguishing between a $\tau=15$ Gyr and 30 Gyr, both producing rather flat SFH, hence leaving a large leverage to the model to converge on both values with similar confidence. On the other hand, the stellar models can be rather insensitive to an $\rm age =0.5$ Gyr, being the broad band photometry unable to catch the typical feature of young stars, and also given the very shallow limiting magnitude of the $u$-band that would provide most of the UV rest-frame emission of galaxies up to $z=0.9$. However, for this test we decide to maintain a broad range of priors for the parameter space to learn their impact and confidently optimize their choice for future analyses.

As far as the output of both stellar population codes is concerned, {similarly} to stellar masses in Sect. \ref{sec:masses}, star formation rates should also be corrected for the total fluxes. This is needed to ensure that the specific star formation rate, $\rm sSFR=SFR/M_*$, of a galaxy is conserved. Hence, in the following, we will correct the SFRs by the same amount of the stellar masses, i.e. 
\begin{equation}
\rm \log SFR_{corr}=\log SFR_{out} + (\log M_{*,corr}-\log M_{*,out}),
\end{equation}
where $\rm M_{*,out}$ and $\rm SFR_{out}$ are the output of the SED fitting codes and $\rm M_{*,corr}$ is given by Eq. \ref{eq:corr}.

Finally, as we want to select a {star-forming} sample, we will adopt a canonical cut in specific star formation rate (sSFR) to separate passive from active galaxies and use $\log {\rm sSFR}/{\rm yr}^{-1} =-11 $ as a threshold (see e.g., \cite{Ilbert2015,Katsianis2019,Zhao2020,Katsianis2021,Corcho-Caballero2021,Huang2023,Houston2023,Paspaliaris2023}). {SSFRs} lower than this value should not be taken, in principle, at face value as these correspond to a physically negligible SFR. For this reason, we do not use them in our analysis, although we report them in our catalog with the warning to use them with caution. 

\begin{table*}[t]
\footnotesize
\begin{center}
\caption{Star Formation Rates. Summary of the statistical estimator of the different configurations as in Table 
\ref{tab:model_par}.}
    \begin{tabular}{lcccc}
  \hline
   \hline
\\[0.1pt]
\bf Test& \bf LP/BC03/ExD & \bf  LP/M05/ExD & \bf CI/BC03/DelEx & \bf CI/M05/DelEx\\
 & \bf (NE) & \bf (NE) & \bf (NE)  &  \\
\\[0.1pt]
  \hline
\\[0.1pt]
\multicolumn{5}{c}{\bf Spectroscopic Redshifts}\\
\\[0.1pt]
\hline
\\[0.1pt]

Bias & --& -0.343 &  -0.186 &  -0.437 \\
\\[0.1pt]
NMAD & --&  0.302 &  0.292 &  0.303 \\
\\[0.1pt]

Out. frac. & -- &  4.9\% &  5.6\%  & 5.0\% \\
\\[0.1pt]
  \hline
\\[0.1pt]
\multicolumn{5}{c}{\bf Morphoto-metric Redshifts}\\
\\[0.1pt]
\hline
\\[0.1pt]
Bias &  0.104 &  -0.258 &  -0.098 &  -0.366 \\
 &  (0.090) &  (-0.273) &  (-0.090)  & -- \\
\\[0.1pt]
NMAD &  0.163 &  0.330&   0.324 &  0.346  \\
 &  (0.141) &  (0.335) &  (0.338)  & -- \\
\\[0.1pt]

Out. frac. &   4.5\% &  5.0\%  & 4.9\% &  4.7\% \\
& (4.8\%) &  (5.0\%)  & (4.8\%) &   -- \\
\\[0.1pt]
  \hline
   \hline
    \end{tabular}
    \label{tab:sfr_stats}
    
\end{center}
\textsc{Note.} --- 
Statistical properties of the different model combinations from Table \ref{tab:model_par} as referred to the LP/BC03/ExD from the spectroscopic redshifts. Results for the Nebular Emissions (NE) are given between brackets for the models using the GaZNets redshifts as input.
\end{table*}

\subsubsection{Using Spectroscopic redshifts}
As for the stellar masses in Sect. \ref{sec:masses}, we first discuss the SFR results obtained using the spectroscopic redshift as fixed parameter in 
LePhare and CIGALE.
In Fig. \ref{fig:B03vsall_SFR}, we show the SFRs computed using the different libraries and SFHs as in Table \ref{tab:model_par}. Overall the SFRs look all aligned along the 1-to-1 relation, although both LePhare and CIGALE estimates using M05 show some negative offset (more pronounced in CIGALE), as seen by the residuals shown at the bottom of each panel.
Furthermore, at $\log {\rm SFR}/M_\odot {\rm Gyr}^{-1}\lsim 8$, the correlations show a tilt toward a positive bias, more pronounced for CIGALE, that only for CI/BC03/DelEx partially compensates the negative bias at higher SFRs.

On the other hand, at $\log {\rm SFR}/M_\odot {\rm Gyr}^{-1} \gsim 8$, the CI/BC03/DelEx estimates are nicely consistent with the LePhare estimates of LP/BC03/ExD. 
Overall the two tools show a substantial agreement if they use the same libraries, while {they} do not seem to show a strong dependence on the SFH.
This is seen from Table \ref{tab:sfr_stats}, showing the statistical estimators for the different experiments. Here we find, indeed,
that LP/M05/ExD and CI/M05/DelEx have similar Bias, NMAD, and outlier fraction.

Looking at the whole statistical estimators 
we can confirm that, broadly speaking, the relative bias of the SFR estimates is 
barely consistent with zero within the NMAD for M05, while it is well consistent with zero for the BC03. From Fig. \ref{fig:B03vsall_SFR} (bottom panels), we also see that the overall scatter of the residual is of the order of 0.3 dex, slightly larger to the one of the stellar masses.
Moreover, the outlier fraction, even in this case is consistent with a log-normal distribution, across the all SFR range. This broad result suggests that the SFR in star-forming galaxies is a rather stable parameter, in the redshift range we have considered. The degree of accuracy and scatter among different model and library configurations is almost comparable to the stellar masses derived from spec-$z$. We will now check, if this works similarly for the morphoto-metric redshifts, while we will check the impact of the NE in Sect. \ref{sec:NE_sfr}.

\begin{figure*}[t]
\centering
    \hspace{0.cm} 
    \includegraphics[width=0.45\linewidth]{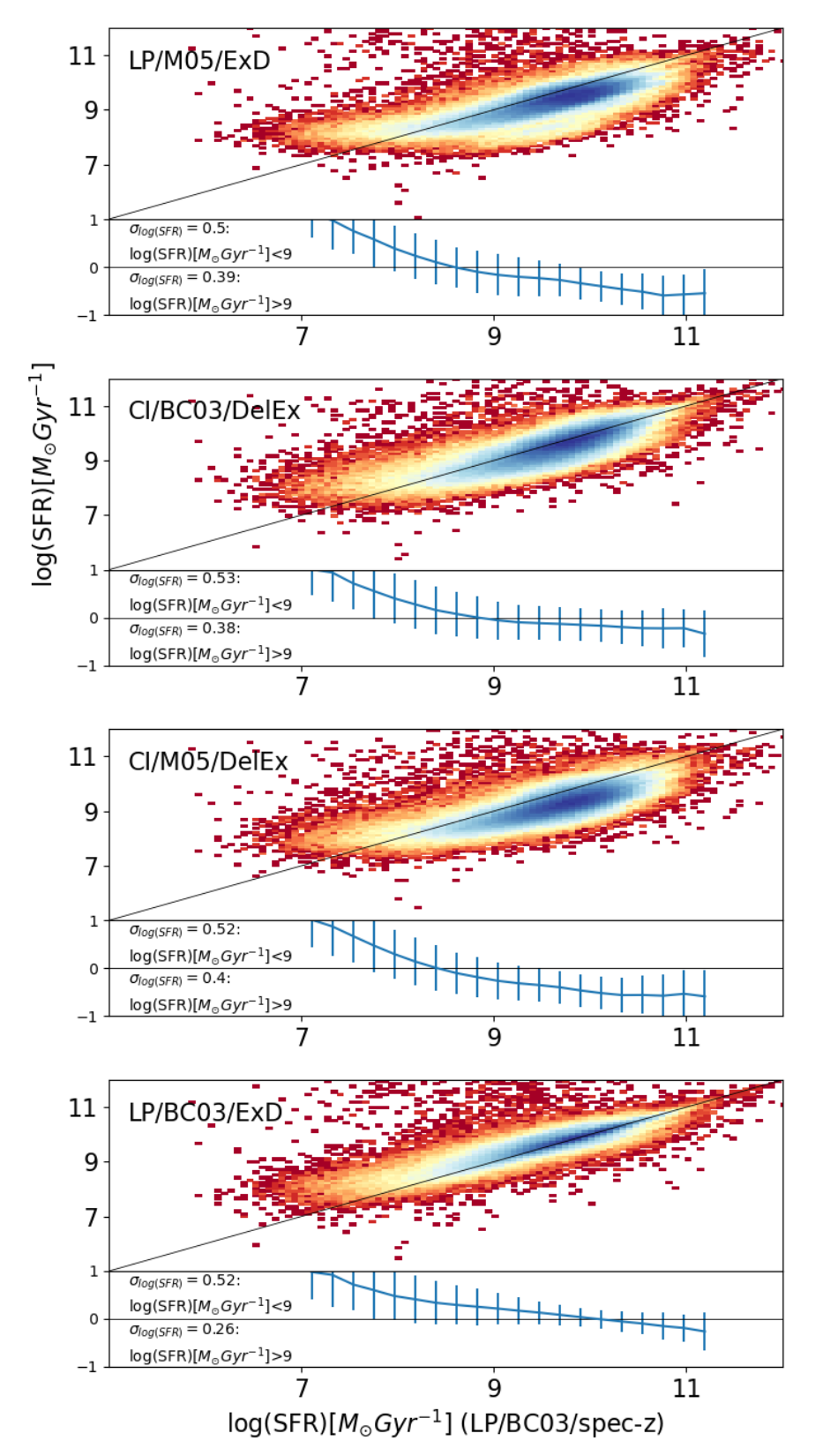}
    
\caption{Star Formation Rates (SFR) estimates as for Fig. \ref{fig:B03vsall_SFR} but using the GaZNet morphoto-metric redshifts. }
\label{fig:B03vsall_GAZ_SFR}
\end{figure*}

\subsubsection{Using morphoto-metric redshifts}
In Sect. \ref{sec:mass_photoz} we have discussed the impact of the morphoto-$z$ from GaZNet on the stellar mass estimates and shown that the net effect of the morphoto-metric redshift is to increase the scatter and the outlier fraction of the final estimates. We have also seen that the overall impact of the GaZNet redshift can be quantified by comparing two set of estimates with same tool, stellar population library and SFH and changing only the input redshits (e.g. using the LP/BC03/ExD from morphoto-$z$ vs. spec-$z$). The same trend is seen for the SFR estimates with the GaZNet redshift with respect to the spec-$z$, as shown in Fig. \ref{fig:B03vsall_GAZ_SFR}. Compared to the spec-$z$ estimates in Fig. \ref{fig:B03vsall_SFR}, we see that the scatter and the number of ``large'' outliers (see below) of the morphoto-$z$ based estimates is increased with respect to the LP/BC03/ExD from spec-$z$. This is seen from the bottom panels with residuals in $\Delta \log \rm SFR$ (see caption), where we report an average scatter of 0.35 dex for $\log {\rm SFR}/M_\odot {\rm Gyr}^{-1}>9$ and 0.52 dex for $\log {\rm SFR}/M_\odot {\rm Gyr}^{-1}<9$.
This is also quantified in Table \ref{tab:sfr_stats}, where we again measure an increased NMAD for all configurations. As noticed for the masses, this is compatible with a pseudo-Gaussian increase of the NMAD values of the morphoto-$z$ estimates, being the NMAD of the LP/BC03/ExD/morphoto-$z$ ($0.163$) a measure of the overall impact of the morphoto-$z$ errors.
The ``gaussianity'' of the $\log \rm SFR$ distribution obtained from the morphoto-$z$ is confirmed by the outlier fraction above the $2\sigma(\log \rm SFR)$, of the order of 5\%.

In the same Fig. \ref{fig:B03vsall_GAZ_SFR}, we also see that the bias is generally compatible with zero, except for the CI/M05/DelEx (morphoto-$z$). In 
general, a trend of the bias with the SFR is evident, due to a positive bias for the lower star formation rates ($\log {\rm SFR}/M_\odot {\rm Gyr}^{-1} \lsim 8.5$). Here the effect of the morphoto-$z$ is to exacerbate the weak trend shown by the spec-$z$ estimates, which is partially absorbed by the scatter of the residuals. 
Due to the well known correlation between the SFR and the stellar mass (see Sect. \ref{sec:sci_val}), we conclude that this has the same origin of the bias found for stellar masses at $\log M_*/M_\odot<8.5$, as discussed in Sect. \ref{sec:masses}. We also notice a cloud of outliers at ${\log \rm SFR}/M_\odot {\rm Gyr}^{-1} \gsim 10$ from the GaZNets-based estimates. These come from a series of morphoto-$z$ outliers, overestimating the intrinsic redshift of the galaxy. Indeed, the higher fictitious redshifts force the SED fitting procedure to interpret the rest frame photometry of the galaxy to be bluer and, hence, more star-forming than the result one obtains from the spec-$z$. 

To conclude the analysis of the SFRs we can say that, as for the stellar masses, these are also a rather stable quantities with respect to the fitting tool, stellar libraries and SFHs, as they do not show significant systematics, except for small SFRs, although we register a tendency of the M05 models to underestimate the SFRs with respect to the BC03.

\subsubsection{The impact of the nebular emissions on star formation rates}
\label{sec:NE_sfr}
We can finally check the impact of the nebular emissions on the predictions of the star formation rates from the different stellar population models considered. As done for the stellar masses in Sect. \ref{sec:NE_masses}, we report the results of the main statistical indicators in Table \ref{tab:sfr_stats}, face-to-face with the same indicators from the no-emission models, using the GaZNet redshifts as input. As for the masses we do not see any significant change on the overall relative bias, NMAD and outlier fraction, meaning that the inclusion of the nebular emissions does not produce any relevant effect for any of the model, given the mass ($\log M_*/M_\odot>8.5$) and redshift range ($z<1$) considered here.

\subsection{Median mass and SFR estimates}
\label{sec:medians}
A relevant result of this paper is that both the stellar mass and the star formation rate are two quantities that can robustly be constrained with seeing matched photometry covering a wide range of wavelengths, from optical to NIR (see e.g. \cite{2010AJ....140.1528Labarbera_optNIR}, \cite{2011MNRAS.418.1587Taylor_gama}). 
For completeness, in \ref{appendixA} we briefly test the case where only optical bands are available and compare this with results obtained in Sect. \ref{sec:results} to briefly illustrate the advantage of adding the NIR to the optical bands, in terms of accuracy and precision of the stellar population estimates.

By robust constraints, here we mean that the $\rm M_*$ and the SFR estimates do not show statistically significant ``relative'' bias if compared to the estimates from other tools, libraries and star formation histories. As seen in Sects. \ref{sec:masses} and \ref{sec:sfr}, this is generally true for all models considered except LP/M05/SB, as this shows a relative bias of the stellar masses which is systematically larger than the scatter of the overall mass estimates (see Table \ref{tab:masses_stats}, and Fig. \ref{fig:B03vsall}). This makes this model an outlier with respect to all other models (see Sect. \ref{sec:masses_spec}) and we decide to exclude it in the following analysis.
As reference estimates we have arbitrarily chosen the LP/B03/ExD model, but this cannot be taken as ground truth. If we assume that the true values of $\rm M_*$ and SFR have to be found within the interval covered by the adopted models, then we can define the ``median'' value as a reasonable estimator of the ground truth of each of them. To deal with the low number of measurements
available to compute the median, we follow the approach of \cite{2015ApJ...801...97Santini_candels} and adopt the Hodges-Lehmann
estimator, defined as the median value of the means in the linear space of all the possible pairs of estimates in the sample:
\begin{equation}\label{eq:mass_med}
\rm M_* ^{\rm MED}= {\rm median} \left ( \frac{M_{*,i}+M_{*,j}}{2} \right ),
\end{equation}  
where the $i$ and $j$ indexes vary over the different models in Table \ref{tab:model_par}. For a dataset with $n$ measurements, the set of all possible two-element subsets, for which the median is computed, has $n(n - 1)/2$ elements. 
Similarly we will define a median star formation rate
\begin{equation}\label{eq:sfr_med}
\rm SFR ^{\rm MED}= {\rm median} \left ( \frac{SFR_i+SFR_j}{2} \right ).
\end{equation}  

Assuming these quantities to be unbiased estimators of the ground truth, we will use them for a science validation test as in the Sect. \ref{sec:discussion}. 
As a sanity check for our median estimates, as well as for individual model results, in Appendix \ref{appendixB} we show a direct comparison of the $M_*$ and SFRs versus some external catalogs overlapping with the KiDS area. In particular, we use the stellar masses from \cite{2011MNRAS.418.1587Taylor_gama}, which makes use of $ugriZ$ photometry, and the SFR estimates of the SDSS-DR7 galaxy sample from the MPA-JHU group\footnote{https://wwwmpa.mpa-garching.mpg.de/SDSS/DR7/sfrs.html \label{cat_sfr_dr7}}, based on spectroscopy data as discussed in \cite{Brinchmann2004_SFR}. The main conclusion from these comparisons is that the ``median'' stellar masses and the star formation rates derived from the 9-band SED fitting are generally consistent with independent estimates based on different data and techniques. This is particularly true for $M_*$ estimates, while for SFRs we can expect some offset due to intrinsic systematics of the different proxies adopted (see also Sect. \ref{sec:SFRF}). However, in all cases the relative bias between different datasets is confined in the typical scatter of the data.

\section{Discussion}\label{sec:discussion}
In the previous sections we have assessed the accuracy and scatter of the different configurations using the relative bias, NMAD and outlier fraction as statistical estimators, and concluded that both stellar masses and SFRs are rather robust quantities. In this section we want to  examine the accuracy and scatter in more details, as a function of the intrinsic properties of galaxies, like redshift, signal-to-noise ratio, and stellar mass. This will allow us to check the presence of ``trends'' in the systematics that might affect the stellar population parameters in different volumes of the parameter space defined by these quantities. This is fundamental if one wants to study the evolution of the mass function of scaling relations like the  ``main sequence'' of the star-formation galaxies from the $M_*-\rm SFR$ relation. We also briefly discuss other sources of systematics, and finally compare the galaxy mass function and the $M_*-\rm SFR$ derived from our ``median'' parameters with {previous literature}, as a science validation of our inferences.   

\begin{figure*}
\hspace{-0.6cm}
\includegraphics[width=0.53\linewidth]
{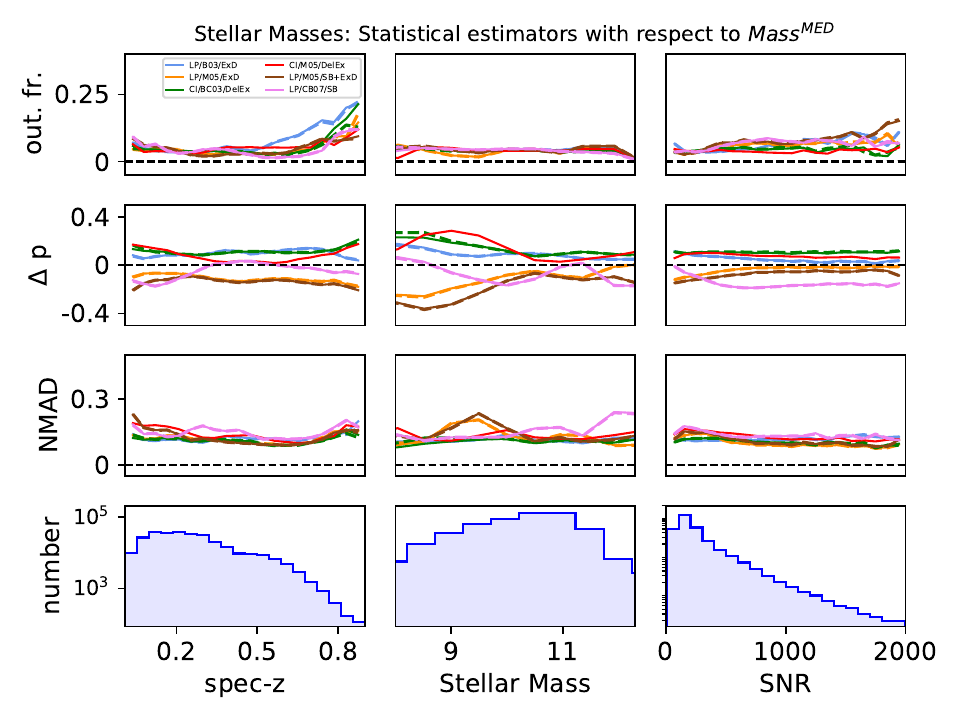}
\hspace{-0.35cm}  
\includegraphics[width=0.53\linewidth]
{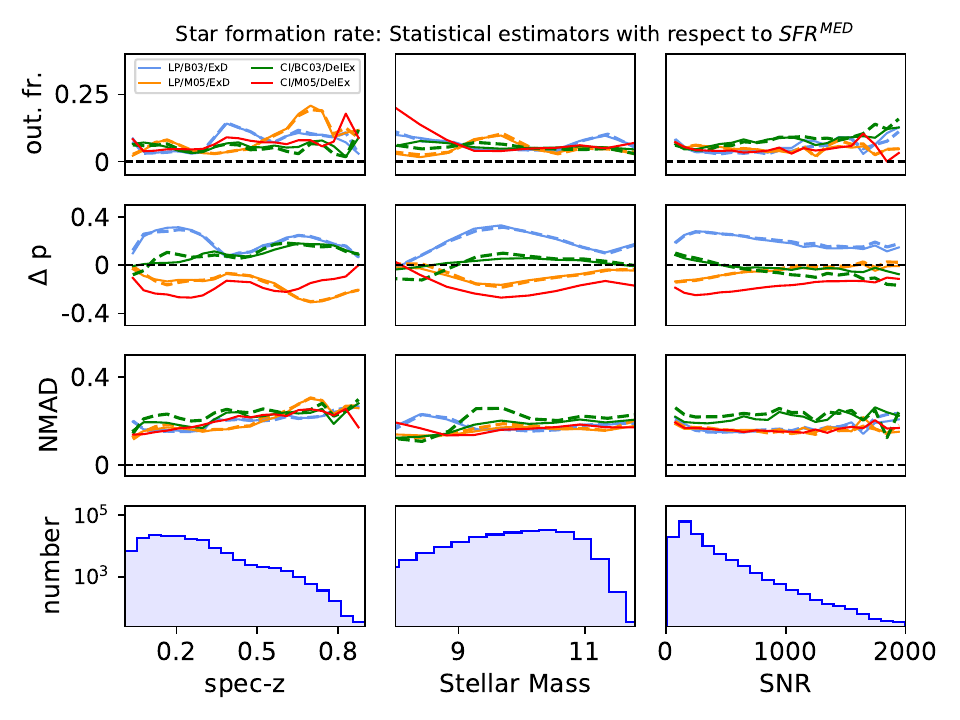}

\caption{Outlier fraction, relative bias and NMAD as a function of redshift, stellar mass and $r$-band magnitude for stellar masses (left) and star formation rates (right) using the morphoto-metric redshift from GaZNet as input. The reference values here are the $M_*^{\rm MED}$ (left) and the $\rm SFR^{\rm MED}$ (right).}
\label{fig:estim_properties}
\end{figure*}

\subsection{Relative bias, NMAD and outliers as a function of redshift, SNR and stellar mass}
\label{sec:estim_galprop}
In Fig. \ref{fig:estim_properties} we plot the bias, NMAD and outlier fraction as a function of the redshift, $r$-band signal-to-noise ratio (SNR) and stellar mass, for the stellar masses (left) and star formation rates (right) derived fixing the redshifts to the morphoto-$z$. Here, we decide to show only the GaZNet-based estimates because, as seen in Tables \ref{tab:masses_stats} and \ref{tab:sfr_stats}, these are the estimates that, by incorporating the uncertainties of the morphoto-metric redshifts, provide the upper limits for both scatter and outlier fractions. 
We also show the dependence on the $r$-band SNR as lower limit of the photometry uncertainties (being all other bands generally less deep than the $r$-band), that should also enter in the precision of the stellar population parameters. We finally remark that we limit our comparison to $\log M_*/M_\odot>8.5$, as we have seen in Sect. \ref{sec:masses} that, below these limit masses, the estimates are dominated by the morphoto-$z$ biases.

The first comment is that both stellar masses and star formation rates show similar features of the statistical estimators as a function of the different quantities, suggesting that the source of the biases and scatter are the same for both quantities. For stellar masses, starting from the top to the bottom, the outlier fractions stay usually within more than acceptable values at all ranges, although we see that toward low redshift ($z\lsim0.05$) and masses (\rm $\log M_*/M_\odot\lsim9$) the outlier fraction and NMAD show a systematic increase. This has been anticipated in Sects. \ref{sec:masses} and \ref{sec:sfr} and tracked to an excess of outliers on the GaZNet redshifts. 
Similar degradation of the estimators are observed at $z\gsim0.7$ for $\rm M_*$ 
estimates, mainly for the poorer statistical samples, which have also degraded the redshift estimates. Overall we see that all the statistical estimators remain contained within reasonable bias ($\rm |\Delta p|<0.2$), NMAD ($<0.3$) and outlier fraction ($<10\%$), especially 
having excluded the LP/M05/SB from the mass set-ups. For star formation rates, notice 
the bimodal {behavior} of the $\rm |\Delta p|$ between the M05 and the BC03 models discussed in Sect. \ref{sec:sfr}, with the 
{CI/M05/DelEx} model showing the largest deviation from all other models. This possibly suggests that M05 stellar libraries need to cope with more complex SFHs than the ones used here. However, the overall indicators of all models seem to stay contained {in} the limits of the NMAD, we have kept all of them in the $\rm SFR^{MED}$ estimates, to average off all possible systematics.
For all the other estimators (NMAD and outlier fraction), we see little difference among the adopted fitting configurations, and confirm no major impact of the NE models either. 
To conclude, we expect we can use the ``median'' estimates in the full range of masses $\log M_*/M_\odot>8.5$ and at all redshift $\lsim 1$ in future applications, although, for SFRs, it remains to see if the $\rm SFR^{\rm MED}$ is totally bias free. In the next sections, and in Appendix \ref{appendixB}, we will show some evidence that this might be the case. We have also checked the statistical estimators as a function of $r$-mag (not shown) and we can confirm that the outlier fraction and the bias become almost out of control at $r$-mag $\gsim23$, which sets a safe limit for future applications based on the use of current GaZNet redshifts.

\subsection{Some considerations about other sources of systematics}
Before moving to some science application we need to stress that providing a full insight {into} all possible systematics that might come from the stellar population analysis is beyond the purpose of this paper. 

We have already introduced the problem of the wavelength coverage in Sect. \ref{sec:medians} and addressed in \ref{appendixA}. 
Other sources of bias one should consider are the input redshifts in the stellar population tools. As we have discussed in Sect. \ref{sec:photo_spec}), we are motivated to do this because by leaving the stellar population tools to {constrain} redshift and stellar population properties at the same time, we expect the degeneracies between redshifts and galaxy colours to strongly affect the stellar populations. This is also briefly discussed in \ref{appendixC}, where we show that the results in terms of photo-$z$ and stellar masses are much more scattered and prone to biases than fixing the redshifts. 
On the other hand, we have seen in the previous sections that in case of unbiased morpho/photometric redshifts moving from spectroscopy to photometry based redshifts, the accuracy is not affected, but the scatter and the outlier fraction is increased with an acceptable level.

Finally, a comment on the stellar templates. In this paper we have used a variety of libraries that could be directly incorporated in the two reference tools adopted (see Table \ref{tab:model_par}). However, this list is neither complete, nor optimal to really account for the current state-of-art stellar population models. We expect to expand our analysis to other stellar libraries (see, e.g., MILES, \cite{2010MNRAS.404.1639Vazdek_MILES}), in future analyses. In this respect, we can consider this analysis as a first step to a more general program to implement a larger variety of models to ground based multi-band datasets.

\begin{figure*}
    \begin{center}
    \includegraphics[width=0.48\linewidth]{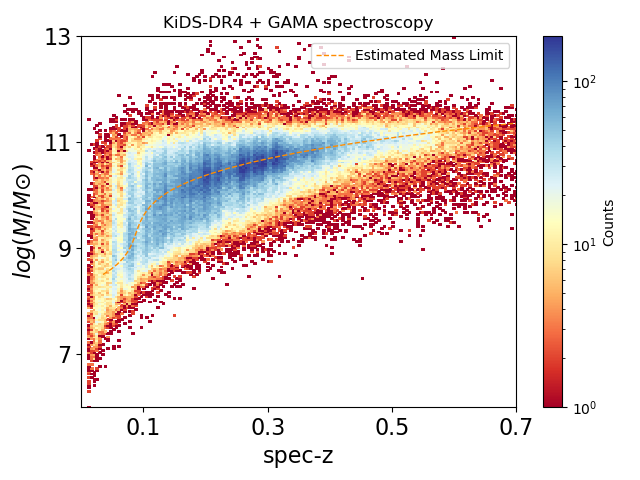}
    \includegraphics[width=0.48\linewidth]{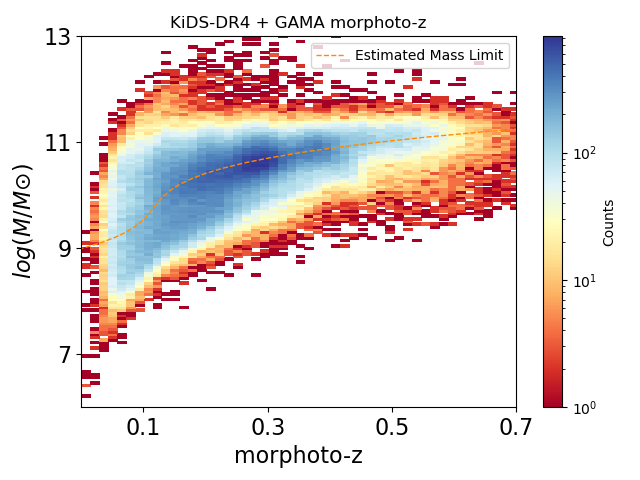}
 
     \end{center}
   
    \caption{Median stellar masses vs. redshift for the fixed spec-$z$ (left) and morphoto-$z$ (right). The dashed line shows the local mass completenes derived in narrow redshift bins and interpolated.
    }
    \label{fig:mass-z}
\end{figure*}

\subsection{Galaxy stellar mass function, star formation rate function} and SFR-$M_*$ relation \label{sec:sci_val}
We want to conclude this paper with a science validation test for the quantities we have focused on in this analysis: the ``median'' values, $\rm M_*^{MED}$ and $\rm SFR^{MED}$. In Sect. \ref{sec:medians} we have seen that these quantities can be considered robust estimates of the stellar mass, $\rm M_*$, and the star formation rate, $\rm SFR$, respectively. A way to test this, is to derive the galaxy stellar mass function (GSMF), i.e. the number of galaxies in a given mass bin in a unit of volume, $\Phi\rm (M)$, and the corresponding star formation rate function (SFRF), i.e. the number of galaxies in a given SFR bin in a unit of volume, $\Phi\rm (SFR)$. This latter, in particular, will give us the chance to compare the SFRs derived from different indicators (UV, H$\alpha$, IR luminosities) with our estimates obtained from the KiDS 9-band photometry. We finally derive the $\rm SFR-M_*$ relation and compare them with independent observations to check for broad consistency of our inferences with previous {literature}. This will allow us to qualify the dataset based on the process presented in Sect. \ref{sec:results} for future catalog compilations and science applications. 

Both the GSMF and the $\rm M_*-SFR$ relation have a crucial role in the understanding of the assembly and formation of galaxies (see discussion in Sect. \ref{sec:intro}) and there have been enormous progresses to trace these quantities back to the early phases of the galaxy formation (see e.g., GSMF: \cite{CanoD2019}, \cite{driver_2022_gama_dr4}, \cite{2022arXiv221202512W_cosmos2020}, \cite{Lovell2022}, \cite{Kim2022}, $\rm SFR-M_*$: \cite{Whitaker2014,Donnari2019,Davies2019,Katsianis2020,Leja2022,Sandles2022,Popesso2023,Kouroumpatzakis2023}). The SFRF is less constrained (especially for high {star-forming} galaxies) as it is highly dependent on the assumed methodology to obtain the galaxy SFRs (see e.g. \cite{Katsianis2021}).

For this test, we are interested to check the consistency of our derivations with previous literature in a statistical sense, while we leave the physical interpretation of these relations for a dedicated analysis, using the full KiDS photometric galaxy catalog. To avoid corrections due to the different completeness mass of the GAMA and SDSS-DR17 data in our spectroscopic sample (see Sect. \ref{sec:photo_spec}), we will consider, here below, only the GAMA sub-sample.

\subsubsection{Galaxy Stellar Mass Function}
\label{sec:gsmf}
In Fig. \ref{fig:mass-z} we start by showing the stellar mass vs. redshift diagram of the GAMA galaxies in our sample. We also overplot the contour of the completeness mass, obtained from the turn-over points in number counts in a given (narrow) redshift bin (see e.g. \cite{2019A&A...632A..34Wright_KV450} for more details on this method). As we can see, the completeness mass becomes almost constant to $10^{11} M_\odot$ at $z\gsim0.4$, leaving there just a small statistical 
samples to compare with literature. We then decide to limit our analysis at $z<0.4$, where we have different reference works to compare our data with.

\begin{figure*}
\centering
\includegraphics[width=0.9\linewidth]{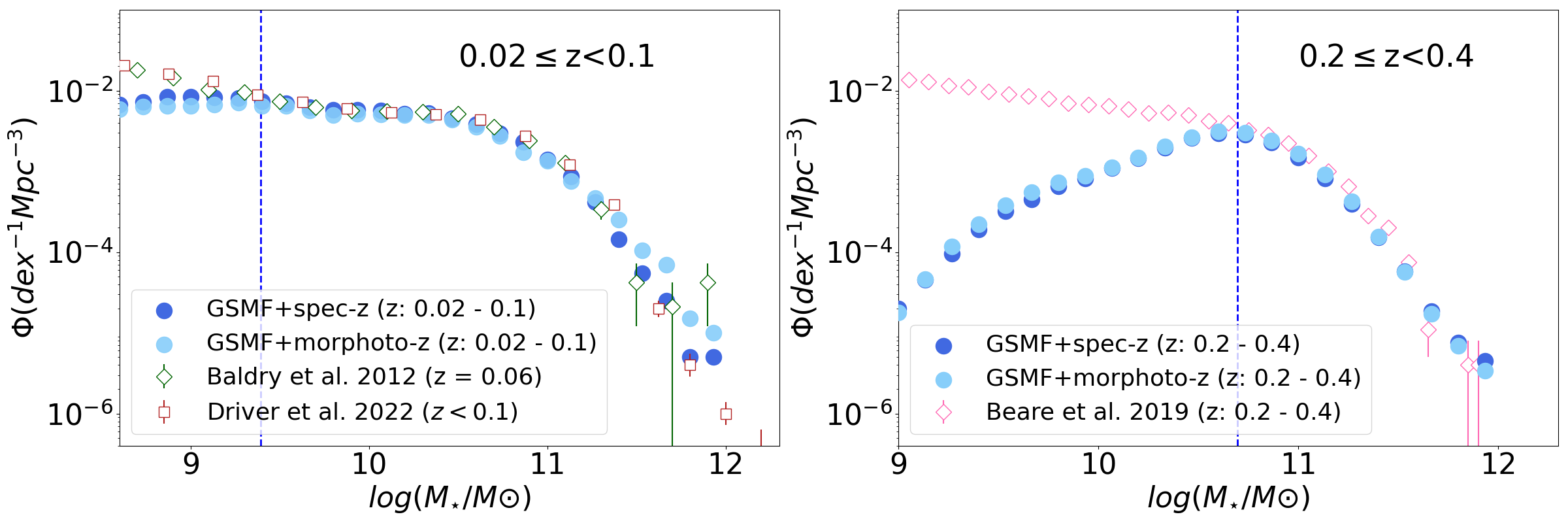}
    \caption{Galaxy Mass Function (GSMF) of KiDS galaxies with GAMA spectroscopy and GaZNet morphoto-$z$. The GSMF is obtained with the median stellar masses, $\rm M_*^{MED}$. Left: the GSMF from the KiDS/GAMA sample in the redshift bin $z={0.02-0.1}$; right: the GSMF in the redshift bin $z={0.2-0.3}$. The reference literature for comparison is given in the legenda (see text for details). The vertical dashed line shows the local mass completeness in each bin.
    }
    \label{fig:GSMF}
\end{figure*}

For the comparison of the GSMF, we use observations derived for the GAMA galaxies at $z<0.1$ (\cite{2012MNRAS.421..621Baldry_2012}, \cite{driver_2022_gama_dr4}) and $0.2<z<0.4$ (\cite{2019ApJ...873...78B_Beare_GMF}).  
In Fig. \ref{fig:GSMF} we show the GSMF from the $M_*^{\rm MED}$ estimates, derived in the redshift bin $z={0.02-0.1}$ and $z={0.2-0.4}$, against the GSMF from similar redshifts for homology. In the same figure, we also show
the completeness mass, defined as in Fig. \ref{fig:mass-z}. 
In Fig. \ref{fig:GSMF}, we do not compute the volume occupied by the complete sample of galaxies in the GAMA area, $V_{\rm max}$, as this would imply to know the GAMA survey selection function, which is beyond the scope of this comparison. We rather normalize the counts to match the literature GSMFs. As we see both the estimates derived by spec-$z$ and the morphoto-$z$ nicely follow the GSMFs of previous literature in the two redshift bins. 

In particular, at z$<0.1$ (left panel) the consistency with previous GAMA inferences from \cite{2012MNRAS.421..621Baldry_2012} and the recent compilation from \cite{driver_2022_gama_dr4} are almost indistinguishable for masses above the limiting mass of our spectroscopic sample, although the match becomes more insecure at very high masses, where both the exact volume adopted and the different selections can cause noisy statistics.  
A similar behaviour is also seen in the other redshift bin adopted ($0.2<z<0.4$, right panel). 
Here, the consistency of our GSMF with the dataset from \cite{2019ApJ...873...78B_Beare_GMF} is yet very  in the full range of masses above the completeness limit.
Overall, this good match with independent GSMFs brings us to the conclusion that the stellar masses we have produced have high science fidelity to be expanded to further analyses. 

\begin{figure*}
    \centering

\includegraphics[width=0.85\linewidth]{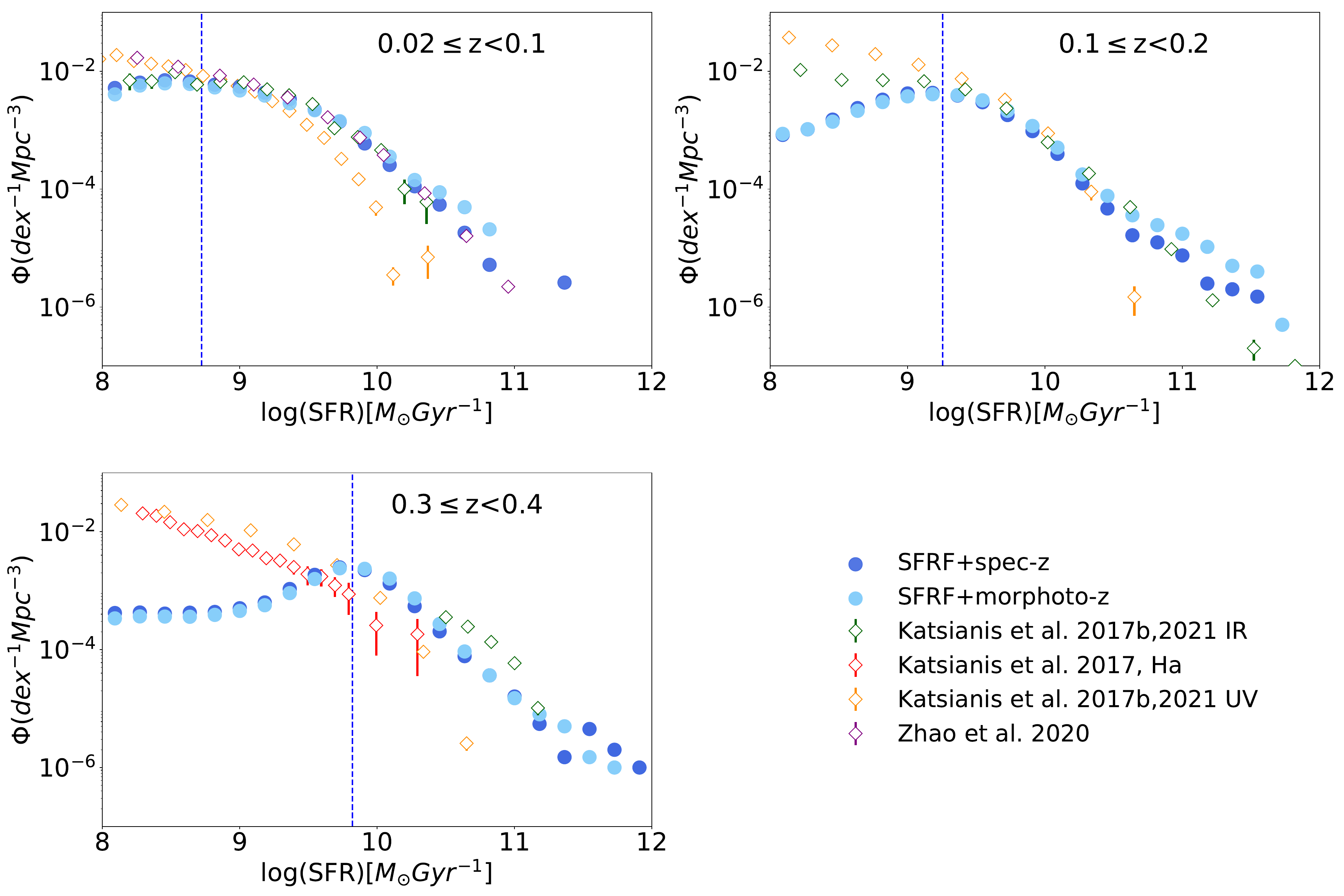}

    \caption{Galaxy Star Formation Rate Function (SFRF) of KiDS galaxies with GAMA spectroscopy and GaZNet morphoto-$z$. The SFRF is obtained using the SFR$^{\rm MED}$ estimates and compared with available litarature in the three different redshift bins: 0.02$\leq z<0.1$ (top-left),   0.1$\leq z<0.2$ (top-right),  0.3$\leq z<0.4$ (bottom-left). Symbols/references are as in the legenda (bottom right). Vertical dashed lines represent the limiting SFR (see text for details). 
    }
    \label{fig:fig_SFRF}
\end{figure*}

\subsubsection{Galaxy Star Formation Rate Function}
\label{sec:SFRF}

Differently from the GSMF, the star formation rate function (SFRF) is not a standard proxy for galaxy evolution, although this can provide relevant insight into galaxy formation (e.g. \cite{Dave2017}). One reason is that SFRs are more sensitive than stellar masses to the assumed methodology. For this reason, more focus is given to the observed values of UV, H$\alpha$ or infrared (IR) luminosity functions as probes of the SFRs in galaxies (\cite{Gunawardhana2013_Ha_LF, Khusanova2020UV+Lya_LF, Wang2021_IR_LF, Donnan2023_UV_LF, Barrufet2023_IR_LF}). Despite these difficulties, there have been attempts to quantify the SFRF at different redshifts (e.g. \cite{Ilbert2015}).
At low redshifts ($z<0.5$) the UV and H$\alpha$ data are significantly affected by dust attenuation effects (e.g. \cite{Sobral2013}, \cite{Gruppioni2016}). This limitation impacts the derived UV/H$\alpha$ star formation rate functions which are usually incomplete at the high star-forming end ({$\log \rm SFR /M_\odot/$Gyr$^{-1}\gsim 10$}).
Thus, especially for these high SFR ranges, 
IR SFRs are considered more robust and give a more accurate estimate of the SFRFs, at least at low redshifts (e.g. \cite{Katsianis2021} and reference therein). Taking all this into account, in Fig. \ref{fig:fig_SFRF} we show the SFRFs based on the ``median'' values derived in Sect. \ref{sec:medians}. We compare these SFRFs in three redshift bins {consistent} with other observations from \cite{Katsianis2017MNRAS,Katsianis2021}, which reports a collection of SFRFs based on UV, H$\alpha$ and IR, and from \cite{Zhao2020}, which presents SFRs from SED fitting of a local sample of SDSS galaxies.

In the figure, we can see the co-existence of SFRs based on different proxies and appreciate the large scatter introduced by the different methods. Broadly speaking, the UV- and H$\alpha$-based SFRFs are consistent between them and generally discrepant from the IR-based ones. Our SED estimates look very well consistent with the IR SFRFs, down to the ``limiting SFR'', marked as a vertical dashed lines in the different redshift bins\footnote{This has been obtained following the same procedure of the stellar masses, i.e. as the peak of the SFRF. Here, though, we do not interpolate in the SFRF vs. $z$ but we show the peak in every particular bin.}. 
Finally, we remark the almost perfect agreement with the SDSS SED estimates from \cite[][see also Appendix \ref{appendixB}]{Zhao2020}, especially considering our spec-$z$ estimates. Hence, we conclude that the SFR$^{\rm MED}$ allow us to build SFRFs which are in good {agreement} with previous literature based on IR luminosity function and SED fitting, while the difference with UV and H$\alpha$-based estimates have to rely to the difference in the calibration of the different methods (see e.g. \cite{Katsianis2020,Katsianis2021}). This does not impact the fidelity of our estimates, as they show no systematics with similar (photometric) probes. As we will see in Appendix \ref{appendixB}, this conclusion is corroborated by the direct comparison of the SFR$^{\rm MED}$ estimates with spectroscopical SFRs, showing a {statistically} insignificant bias for the morphoto-$z$ and no bias for the spec-$z$ based estimates.  

To conclude, the consistency of both the GSMFs and SFRFs with literature further support the {assumption} that the ``median'' estimates represent a realistic proxy of the true $M_*$ and SFRs, either using spectroscopic or morphoto-metric redshifts. In particular, the accuracy of the GMSFs and SFRFs based on morphoto-metric redshifts demonstrate that the method can be successfully extended to larger photometric KiDS galaxy collections.

\begin{figure*}[t]
\centering
\includegraphics[width=0.5\linewidth]{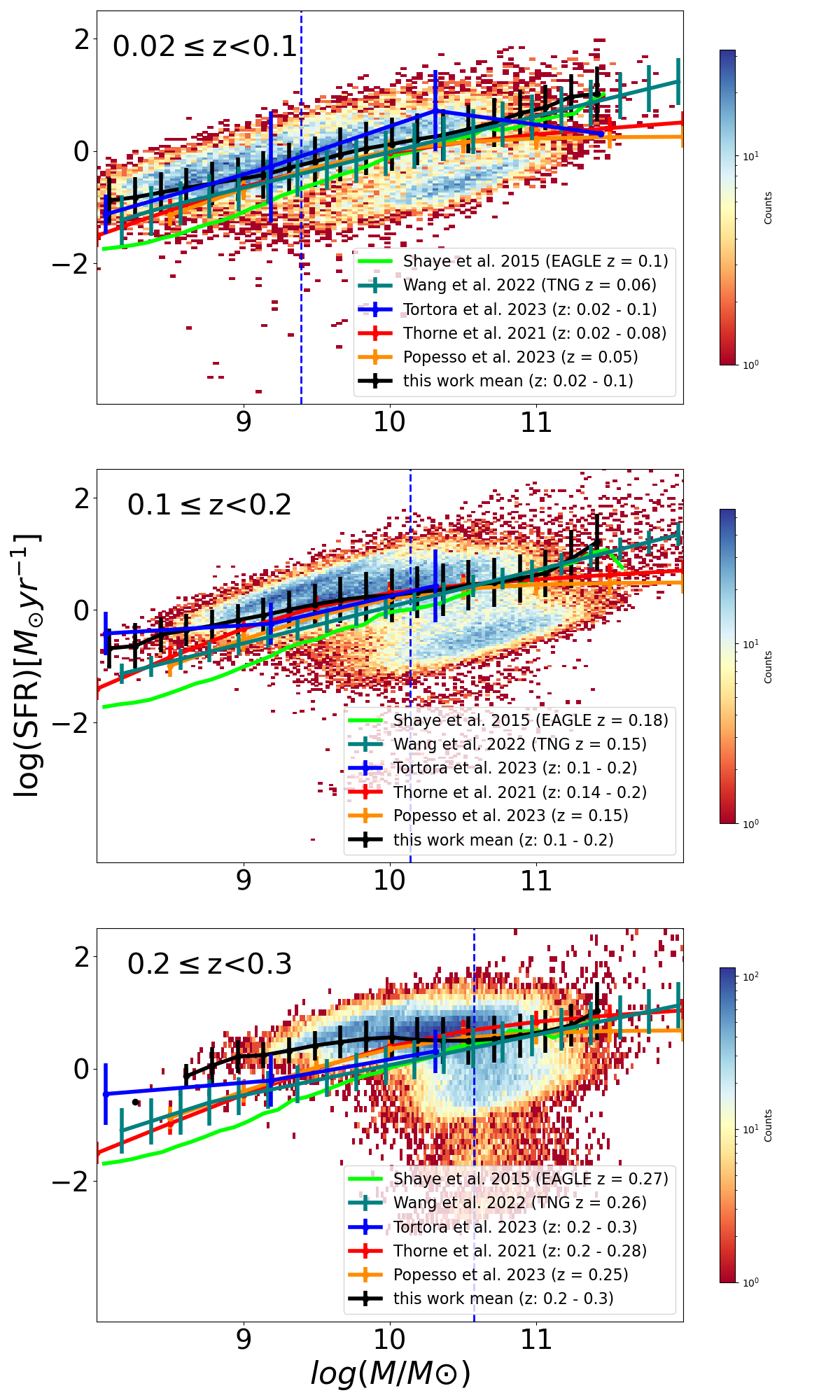}
    \caption{$\rm M_*-SFR$ relation for KiDS galaxies with derived from GaZNet morphoto-$z$. Top: $\rm M_*-SFR$ in the redshift bin $z={0.02-0.1}$; center: in the redshift bin $z={0.1-0.2}$; bottom: in the redshift bin $z={0.2-0.3}$. The reference literature for comparison is given in the legenda (see text for details). The vertical dashed line shows the local mass completeness in each bin.
    }
    \label{fig:mstar-sfr_rel}
\end{figure*}

\subsubsection{$\rm M_*-SFR$ relation}
For the $\rm M_*-SFR$ relation, in Fig. \ref{fig:mstar-sfr_rel} we also plot the results of the lower-redshift bins, where the mass completeness allows us to have 
a sufficient sample for a consistency check. We use, as comparison, 
a series of mean relations of star-forming galaxies from other literature studies in different redshift bins: namely, 1) Tortora et al. (2003, in preparation), including  \cite{Hunt2016a} based on a hybrid method using far-ultraviolet (FUV)+total infrared luminosity, 2) \cite{2021MNRAS.505..540Thorne}, performing SED fitting using {multi-band} FUV-FIR, 3) \cite{Popesso2023}, based on a collection of {homogenized} literature\footnote{They calibrate to a Kroupa IMF, and the SFR estimates to the Kennicutt \& Evans \cite{Kennicutt_Evans_2012}. Note that the choice of IMF does not impact the $M_*-$SFR relation as it equally affects the stellar mass and the SFR estimates.}. 
We also add the prediction from Illustris-TNG (\cite{2022MNRAS.515.3249W}) and EAGLE simulations (\cite{Schaye2015}) to illustrate the potential of deriving SFRs from larger KiDS galaxy samples to check against the outcome of state-of-art hydrodynamical simulations to gain insight on the galaxy formation scenario\footnote{We did not compare the inferred GSMF in Sect. \ref{sec:gsmf} with the same simulations because these latter are tuned, by construction, to fit the observed stellar mass functions.}.

In Fig. \ref{fig:mstar-sfr_rel} we show the $\rm M_*-SFR$ relation for the median quantities obtained using the GaZNet redshifts as input only. This is because 
we have seen, in  
{Sect. \ref{sec:masses}} that these
represent the worse-case scenario, where the measurements are more scattered and 
show systematic effects only at very low masses ($\rm \log M_*/M_\odot<8.5$) -- these are below the completeness mass we can use as lower limit for science analysis. 

From Fig. \ref{fig:mstar-sfr_rel} we find that the $\rm M_*-SFR$ relation of the KiDS galaxies (black points with errorbars) nicely follows the majority of the literature data, both from observations and simulations, down to the completeness mass, despite the different methods adopted in literature and the definition of star-forming systems. At masses below the limiting mass, our $\rm M_*-SFR$ shows a significant departure from other relations.
We expect to check if this is indicative of the presence of systematics, when we will use the full KiDS photometric sample, for which we expect to push the mass completeness to lower levels in all redshift bins.

We are convinced that this consistency check, of both the $\rm M_*-SFR$ relation and the GSMF, which is just qualitative at this stage, confirms the validity of the procedure and the data produced in this analysis. 

\section{Conclusions and perspectives}
\label{sec:concl}
In this paper we have used a spectroscopic galaxy catalog including 9-band ($u g r i Z Y J H K_s$) photometry from the 4th data release of the Kilo-Degree Survey (KiDS) to derive robust stellar masses and star formation histories. We have performed a full template fitting analysis using two popular stellar population codes, LePhare and CIGALE, and a combination of stellar population libraries (\cite{BC03}, \cite{2005MNRAS.362..799Maraston05}, \cite{2007ASPC..374..303BC07}) and star formation histories (i.e. a single burst, an exponential decline, and a delayed exponential). Besides the spectroscopic redshifts, taken from GAMA data releases 2 and 3 and SDSS data release 17, we have considered as input of the SED fitting process, the morphoto-metric redshifts obtained from the deep learning tool GaZNet (\cite{2022A&A...666A..85Gaznet}). In this latter case, we can perform a controlled test of the variance one would introduce, in large dataset, where only photometric redshift are available for the galaxy catalogs. 

In fact, the main goal of this analysis has been to assess the relative accuracy and the variance of the stellar population parameters under a variety of combinations of fitting tools/stellar templates/star formation histories. We summarize here below the main result of this analysis:

1) the stellar mass and the star formation rate show limited scatter and relative bias which is within the scatter, when comparing the estimates for every galaxies against the different methods. As such, these quantities are rather stable against the stellar template fitting set-ups; 

2) the relative bias, NMAD and outlier fraction vary with the stellar mass and SNR, not with redshift;

3) due to the overall resilience of the parameters to the different variables in play, we can reasonably adopt a median definition as an unbiased estimator of the ``ground truth'' values for the parameters. Following \cite{2015ApJ...801...97Santini_candels}, we have used a Hodges-Lehmann median for this robust parameter estimate and used them for a science validation; 

4) we have evaluated the scatter of the individual fitting set-ups with respect to the Hodges-Lehmann median (Fig. \ref{fig:estim_properties}) and found that, depending on the combination of templates and star formation histories, stellar masses and star formation rates can deviate by $\sim$0.1 dex, for high mass systems, to $\sim$0.2 dex, for low mass systems;

5) as a science validation test, we have derived the stellar mass function and the star formation rate mass function, as well as he $\rm M_*-SFR$ {relation} and compared with previous literature in different redshift bins, finding a very good match with a wide literature;

6) we provide the catalog of the galaxy parameters, including stellar masses, star formation rates, age, metallicity, extinction and the $\tau$ of the exponential decaying models, for $\sim290\,000$ galaxies with spectroscopic redshifts, $0.01<z<0.9$, from GAMA and SDSS-DR17. The catalog is available at this URL\footnote{link} and contains also the 9-band GAaP photometry, the $r$-band MAG\_AUTO, and the spectroscopic redshift from the parent spectroscopic surveys.

In the future we plan to expand this test, including more stellar formation tools (e.g. FAST :  \cite{2009ApJ...700..221FAST}, SED3FIT: \cite{Berta2013}, Prospector: \cite{Leja2022}, P12 \cite{Pacifici2023}), star formation histories (e.g. 
log-normal \cite{Abramson2016}, $\Gamma$ \cite{Katsianis2021}) and stellar libraries (e.g. \cite{2010MNRAS.404.1639Vazdek_MILES}). 

This will allow us to investigate an even larger variety of models and use the ``median'' of their outcomes (see Sect. \ref{sec:medians}) as an unbiased stellar population parameter estimators 
for the full KiDS ``galaxy’’ photometric sample and finally provide a general-purpose catalog to be used for a variety of galaxy studies. Piece of similar datasets have been previously used in KiDS, to study the size-mass relation of galaxies (\cite{2018MNRAS.480.1057R}), the ultra-compact massive galaxies number density evolution (\cite{2016MNRAS.457.2845T, Tortora+18_UCMGs, Scognamiglio+20_UCMGs}), the mass function of galaxies at different redshifts (\cite{2019A&A...632A..34Wright_KV450}, the clustering of red-sequence galaxies (\cite{2020arXiv200813154V}), the dark matter halo masses of elliptical galaxies as a function of observational quantities (\cite{2022A&A...662A..55S}), the dark matter assembly in massive galaxies (\cite{Tortora+18_KiDS_DMevol}).

\section*{Acknowledgements}
NRN acknowledges financial support from the Research Fund for International Scholars of the National Science Foundation of China, grant n. 12150710511.
RL acknowledges the support of the National Nature Science Foundation of China (No. 12022306) and the science research grants from the China Manned Space Project (CMS-CSST-2021-A01). 
AK acknowledges financial support from the One hundred top talent program of the Sun Yat-sen University.
HF acknowledges the financial
support of the National Natural Science Foundation of China (grant No. 12203096). 
LX thanks Dr. O. Ilbert for the useful suggestions about LePhare and Fucheng Zhong for useful discussions.
\section*{Data Availability}
The data that support the findings of this study are available at the URLs provided in the text.

\bibliographystyle{unsrt}
\bibliography{myrefs}


\onecolumn

\begin{appendix}

\renewcommand{\thesection}{Appendix}

\section{}

\subsection{\label{appendixA} The impact of missing NIR photometry}

In this Appendix we want to check the impact of the wavelength range on the analysis we have performed, and quantify, in particular, the advantage of the inclusion of the NIR to produce reliable stellar population parameters. It is well known that the wider wavelength base is a necessary prerequisite for accurate photometric redshifts (see e.g. \cite{Hildebrandt2010A&A,Hildebrandt2017MNRAS.465.1454H,2022A&A...666A..85Gaznet}). As we will see in \ref{appendixC}, accurate redshifts have themselves a large impact on the stellar population parameters. 
\begin{table*}
\footnotesize
\begin{center}
\caption{Stellar masses and star formation rates with only $ugri$ photometry.}
    \begin{tabular}{lccccccc}
  \hline
   \hline
\\[0.1pt]
\bf Parameter &\multicolumn{4}{c}{\bf Stellar mass}&\multicolumn{2}{c}{\bf Star formation rate}\\
\\[0.1pt]
   \hline
\\[0.1pt]
\bf Test& \bf LP/B03/ExD & \bf LP/M05/SB &\bf  LP/M05/ExD & \bf LP/CB07/SB & \bf LP/B03/ExD & \bf LP/M05/ExD\\
\\[0.1pt]
  \hline
\\[0.1pt]
\multicolumn{7}{c}{\bf Spectroscopic Redshifts}\\
\\[0.1pt]
\hline
\\[0.1pt]

Bias & 0.020  &   -0.395  & -0.237   &   -0.081 & 0.079 &  -0.156\\
\\[0.1pt]
NMAD  & 0.157  &   0.342  &  0.276    &   0.206 & 0.227 &  0.338\\
\\[0.1pt]
Out. frac. & $4.3\%$ & $3.7\%$ &  $4.2\%$ &  $4.8\%$ & $6.1\%$ & $5.1\%$\\
\\[0.1pt]
  \hline
\\[0.1pt]
\multicolumn{7}{c}{\bf Morphoto-metric Redshifts}\\
\\[0.1pt]
\hline
\\[0.1pt]
Bias & 0.055  & -0.355 &  -0.203 &  -0.043 & 0.151  & -0.108 \\
\\[0.1pt]
NMAD & 0.184  & 0.357 &  0.296 &  0.232 & 0.318  & 0.424\\
\\[0.1pt]
Out. frac. & $3.7\%$ &  $3.7\%$ &  $3.8\%$ &  $4.3\%$ & $5.2\%$ & $5.7\%$
\\[0.1pt]

  \hline
   \hline
    \end{tabular}
    \label{tab:noNIR}
    
\end{center}
\textsc{Note.} --- 
Statistical properties of the different model combinations from Table \ref{tab:model_par} as referred to the LP/BC03/ExD derived for the 9-band photometry and the spec-$z$ (i.e. the reference model as in Tables \ref{tab:masses_stats} and Table \ref{tab:sfr_stats}).
\end{table*}
Here we want to show that, even assuming to correctly know the redshift of a galaxy, the wavelength baseline is crucial to provide stellar masses and SFRs with minimal bias and scatter. For space sake we just consider the extreme case of fully discarding the NIR bands, to show what is the maximum errors one would commit applying the same set-up as in Table \ref{tab:model_par}. For the same brevity reason we show the results for 4 LePhare models: LP/B03/ExD, LP/M05/SB, LP/M05/ExD, LP/CB07/SB. In Table \ref{tab:noNIR} we report the main statistical estimators for the different configurations, for both mass and SFR estimates, either assuming the spec-$z$ or the morphoto-$z$ as input. These can be compared to Tables \ref{tab:masses_stats} and \ref{tab:sfr_stats}. 
The most evident effect is the large increase of the scatter of the estimates, as measured from the NMAD.
For stellar masses we find that the NMAD increases by 30-40\% (e.g. LP/M05/ExD/spec-$z$) to about 100\% (LP/B03/ExD/morphoto-$z$). On the other hand, all SB model show little increase in NMAD ($\lsim 10\%$) and smaller relative biases, indicating that these are almost insensitive to the wider wavelength baseline.
For the SFRs we find a similar degradation of the precision of the estimates with NMAD in Table \ref{tab:noNIR} increased by 30\% to 90\% with respect to Table \ref{tab:sfr_stats}, and minimal variation on the relative bias.

\begin{figure*}[t]
    \centering
    \includegraphics[width=0.75\linewidth] 
    {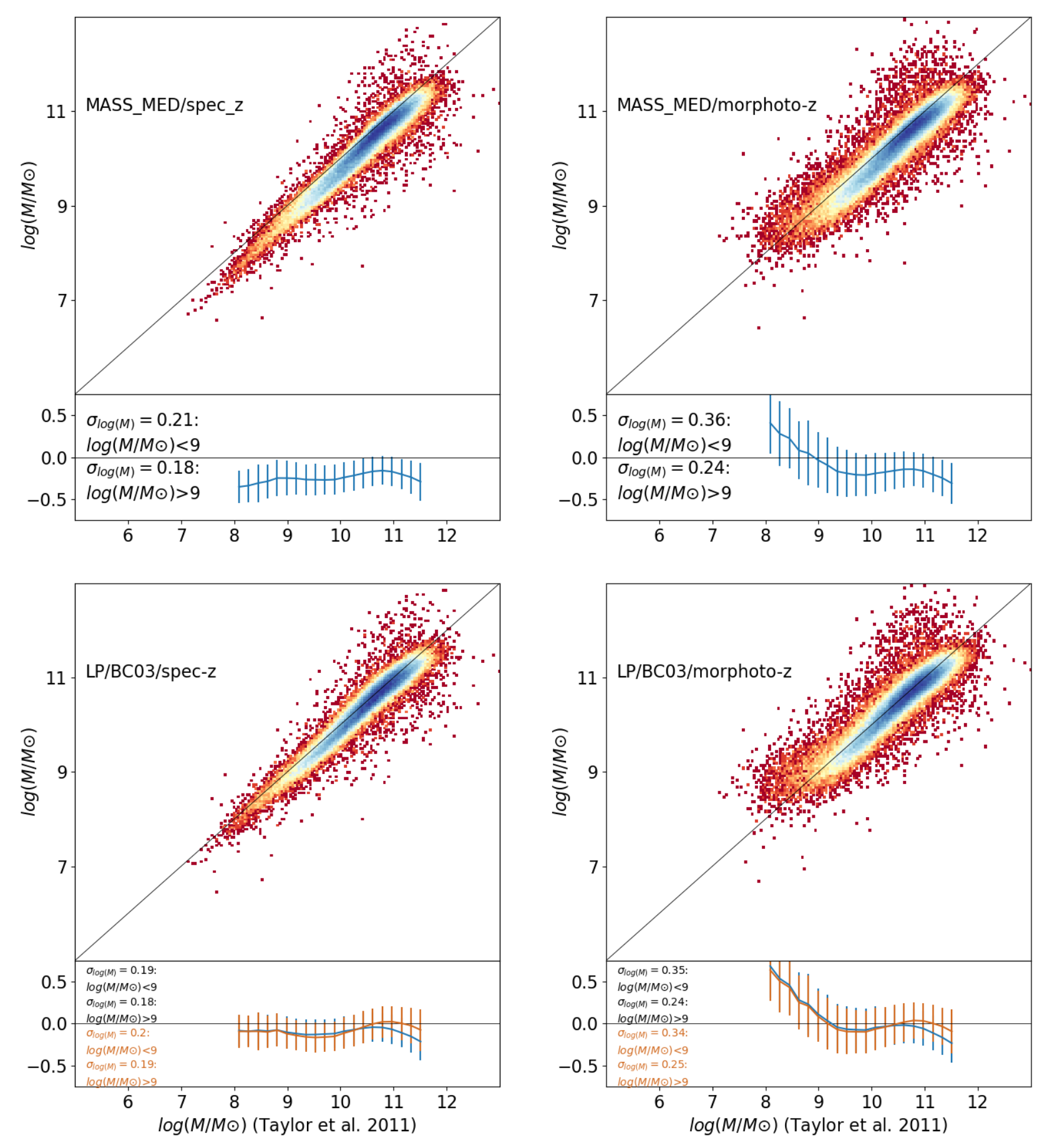}

    \caption{Mass comparison with \cite{2011MNRAS.418.1587Taylor_gama}. Top row: the comparison of the $M_*^{\rm MED}$ the KiDS sample (y-axis) vs. the T+11 sample, using the spec-$z$ (left panel) and morphoto-$z$ (right panel) as input. Residuals, defined as $\Delta \log M=\log M_y-\log M_x$, with respect to the 1-to-1 relation, and the related scatter, defined as the standard deviation, $\sigma(\Delta\log M)$, are shown at the bottom of each panel. Bottom row: same as top row but comparing the LP/BC03 model, which is closer to the one used by T+11. Here we also show the residuals and average scatter using the same band coverage as T+11 ($ugriZ$), in orange.
    \label{fig:mass_compare_Taylor2011}
    }
    \label{fig:taylor11}
\end{figure*}

\subsection{\label{appendixB} Comparison of $M_*$ ans SFR estimates against external catalogs}
As anticipated in Sect. \ref{sec:medians}, here we want to perform a direct comparison of our $M_*$ and SFR estimates against external catalogs. 

For the stellar masses, we have mentioned the existence of stellar mass catalogs based on similar KiDS data (e.g. \cite{2019A&A...632A..34Wright_KV450}, \cite{Bilicki2021A&A...653A..82B}), however here we decide to compare the stellar masses with a catalog made on a different photometric data from \cite[][T+11 hereafter]{2011MNRAS.418.1587Taylor_gama}. The catalog of their stellar masses is available on the GAMA website\footnote{Catalog link: http://www.gama-survey.org/dr2/schema/table.php?id=179}. This is based on the $ugri$ optical imaging from SDSS (DR7) and (according to the catalog description) $Z$-band from UKIDSS (see T+11 and reference therein). Similarly to us they also use BC03 templates, Chabrier IMF, and Calzetti extinction law, with an Exponentially declining star formation history, but they use a customized code for their stellar population models. Hence we can expect some differences in the estimates due to the code adopted and the data (different observations, photometric accuracy and errors etc.), while they use the GAMA spectroscopic redshifts information as input of their model. We have found a match of 64\,771 galaxies with our catalog which are plotted in Fig. \ref{fig:taylor11} against the $M_*^{\rm MED}$ estimates from Sect. \ref{sec:medians}, considering both the spectroscopic and the GaZNet redshifts as input. Being our LP/BC03/ExD the closest model to their set-up, we also add this for comparison in the same figure. Overall, we see that all the estimates (except the $M_*^{\rm MED}$/spec-$z$) are consistent within the errors, shown at the bottom of each panel, with a scatter that is always contained within $\sim 0.2$dex for the spectroscopic redshifts and $\sim 0.25$dex for the morphoto-metric redshifts. We also clearly observe that the LP/BC03/ExD has almost no bias, meaning that the different codes and also the different data have a minimal impact on the final mass estimates. The offset with the $M_*^{\rm MED}$ (of the order of 0.15 dex) is due to the relative bias of the different models entering in the ``median'' quantities: in Sect. \ref{sec:estim_galprop} this is quantified to be $\sim$0.10 dex for the LP/BC03/ExD (see blue line in the 2nd row from the top of Fig. \ref{fig:estim_properties}-- left panel), hence consistent with the 0.15 dex offset above, considering the scatter of $\sim0.2$ dex in the top/left panel of Fig. \ref{fig:taylor11}. The bias with the LP/BC03/ExD model become even smaller is we use the same baseline as T+11, i.e. the 5-bands $ugriZ$, as shown by the orange residuals at the bottom of each panels. This also indicates that most of the effect of the NIR bands mainly impacts the massive galaxies where the difference of the masses can be as large as 0.2 dex. We still see, in all cases, the systematic deviation of the sample based on GaZNet redshifts at $log M_*/M_\odot<9$ discussed extensively before, depending on the redshift systematics and not on the stellar population analysis. 
\begin{figure*}
    \centering
    \includegraphics[width=1.0\linewidth] 
    {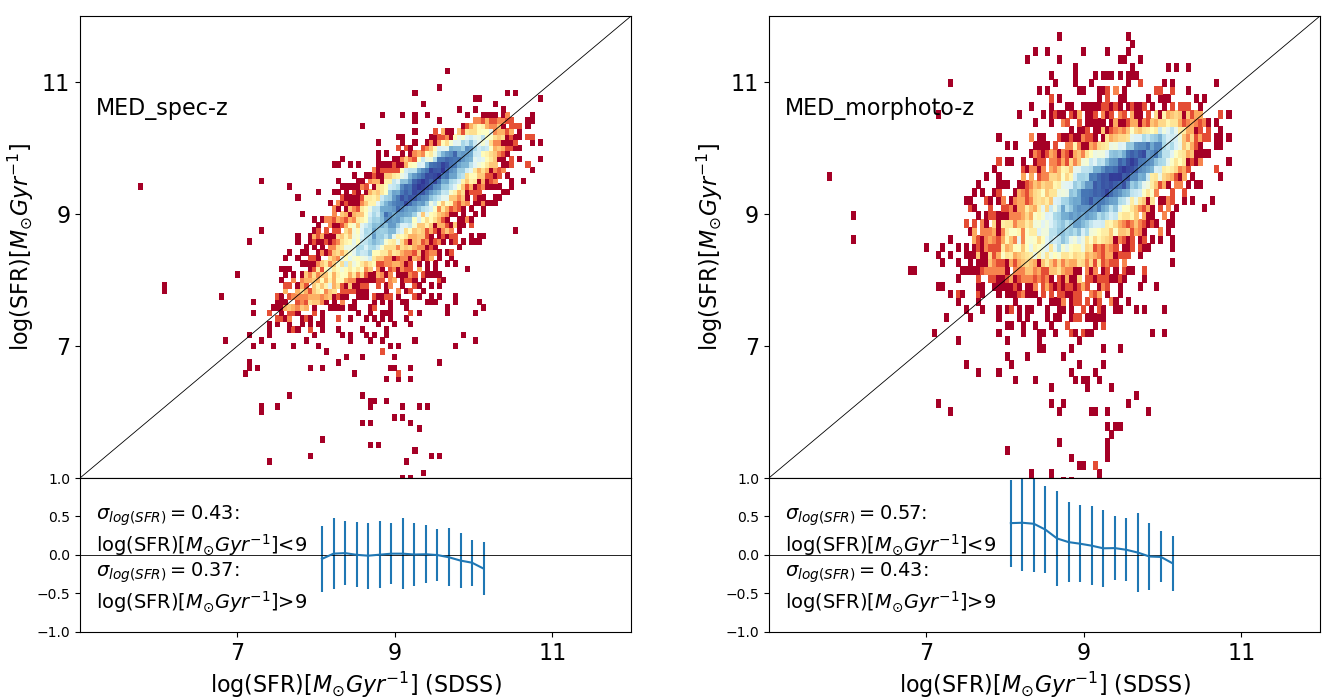}
    \caption{SFR comparison with the SDSS-DR7 sample. Here we use only the median SFR$^{\rm MED}$ along the y-axis for the the spec-$z$ (left panel) and morphoto-$z$ (right panel) as input. Residuals and scatter are as in the top panel of Fig. \ref{fig:taylor11}}.
    \label{fig:SFR_compare_SDSS}
    
    \label{fig:sfr_sdss}
\end{figure*}

For the SFRs we make use of the SDSS-DR7 star formation rate catalog (see footnote \ref{cat_sfr_dr7}) based on the analysis discussed in \cite{Brinchmann2004_SFR}, but see also \cite{Zhao2020}. Here, the star formation rates are computed by directly fitting the emission lines (e.g., H$\alpha$, H$\beta$, [OIII]@$\lambda$5007, [NII]@$\lambda$6584, [OII]@$\lambda$3727, and [SII]@$\lambda$6716). This offers us the opportunity to check the presence of biases on our ``median'' results against spectroscopic inferences, hence based on a more robust method, especially considering the bimodal bias from M05 and BC03, discussed in Sect. \ref{sec:estim_galprop}. The comparison of our 9-band estimates with no-NE and the SDSS-DR7 SFRs are shown in Fig. \ref{fig:sfr_sdss}. We decide to use the no-NE to confirm the little impact of the emission line in the SED based SFR estimated, as discussed in Sect. \ref{sec:NE_sfr}. In Fig. \ref{fig:sfr_sdss}, we see that the SFR$^{\rm MED}$ are in very good agreement with the SDSS spectroscopic inferences, with a bias which is well within the scatter of the data-points. For the morphoto-$z$ sample we see, as usual, the positive bias at low star formation rates induced by the systematics in the morphoto-metric redshifts at low-SFR values, as discussed in Sect. \ref{sec:sfr}, although here the offset starts to become significan at ${\log \rm SFR}/M_\odot {\rm Gyr}^{-1} \lsim 9$, suggesting that the different methods (e.g. emission lines vs. SED fitting) can introduce some biases (see also Sect. \ref{sec:SFRF}). The scatter remains always confined within $\sim 0.4$ dex (see values in the figure insets), in line with the results also discussed in the same Sect. \ref{sec:sfr}, at least for higher SFRs ($\log {\rm SFR}/M_\odot \rm Gyr^{-1}>9$). 

We believe the evidences collected in this appendix, for both stellar masses and SFRs, support all the main conclusions of the paper about {\it the robustness of the stellar population quantities from the different methods/set-ups and the use of the ``median'' values as an unbiased estimator of the true quantities for our galaxy sample}, given the range of redshifts adopted.

\subsection{\label{appendixC} LePhare results with redshift as free parameter}
\begin{figure*}[t]
    \centering
    \includegraphics[width=1.0\linewidth] 
    {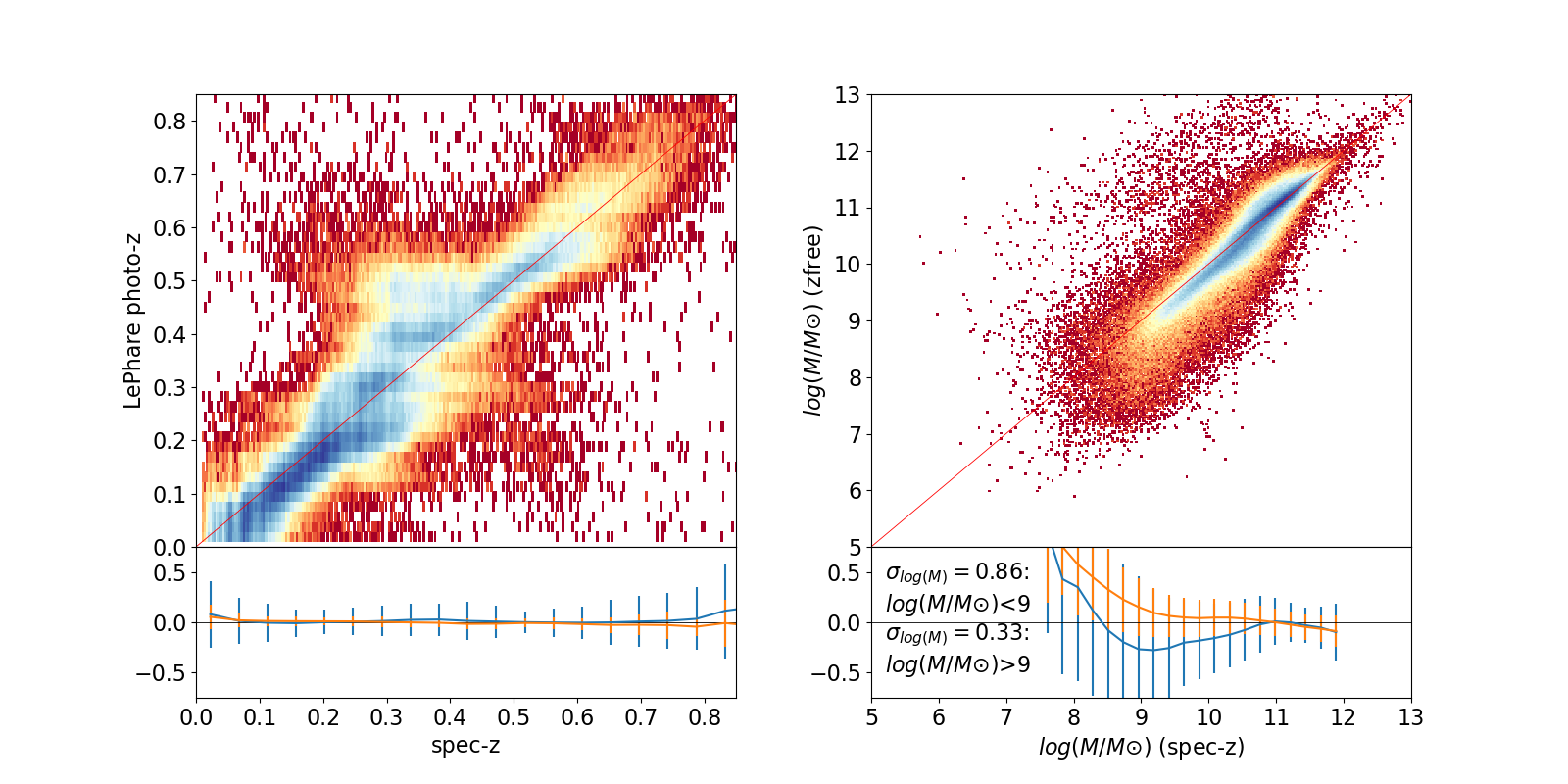}
    \caption{Redshift and Stellar mass estimates from LePhare with redshift as free parameter. Left: the estimated photometric redshift for the configuration LP/BC03/ExD vs. the spectroscopic redshifts. Right: stellar mass estimates using the photo-$z$ as in the left panel vs the same estimates using the spec-$z$. Residuals and scatter with respect to the 1-to-1 relations are reported at the bottom of each panel for the photo-$z_{\rm LP}$ (blue) and GaZNet (orange), as discussed in the text}.
    \label{fig:app_z}
    
    \label{fig:app_z}
\end{figure*}

Both SED fitting tools, LePhare and CIGALE, can use the redshift as free parameter during the fitting procedure. This gives us the chance to have a direct visualization of the degeneracies in the final results, introduced by missing the information about accurate galaxy redshifts. For this particular test we use LePhare to show the impact on the estimates of the stellar masses. We use the reference set-up, i.e. the LP/BC03/ExD, which becomes LP/BC03/ExD/specz for the case with spec-$z$ fixed and LP/BC03/ExD/zfree in the variation with the redshift as free paramater. In Fig. \ref{fig:app_z} we show: 1) on the left panel the spec-$z$ vs the photometric redshift inferred by LePhare, photo-$z_{\rm LP}$, and 2) on the right, the corresponding stellar masses.

From this figure, we can clearly see the impact of missing the information on redshift in the stellar population analysis, in comparison with the equivalent quantities obtained for the GaZNet morphoto-$z$ (Fig. \ref{fig:photo_spec_z} and Fig. \ref{fig:B03vsall_GAZ}, bottom left). This is also quantifies in the residual plots at the bottom of Fig. \ref{fig:app_z}, where we plot the relative bias and scatter both for the photo-$z_{\rm LP}$ and the GaZNet redshifts inferences. In particular, the stellar masses in the former case show a bias and scatter that is fully driven by the larger variance of the photometric redshits: see, e.g., the cloud of galaxies with masses almost parallel to the 1-to-1 relation with a large positive offset on the top of the figure, which are absent in the GaZNet-based estimates. This is confirmed by the global statistical estimators, as for the redshifts we have $\mu=0.005$, NMAD$=0.039$ and Out. frac.$=2.3\%$, i.e. much larger than the same quantities derived for the GaZNet redshift in Fig. \ref{fig:photo_spec_z} ($\mu=0.005$, NMAD$=0.017$ and Out. frac.$=0.4\%$, respectively) which are mirrored by a similar worsening of the same estimators from the masses that for the $z_{\rm LP}$  case are $\Delta p=-0.077$, NMAD$=0.197$ and Out. frac.$=4.7\%$ which are up to twice as much larger than typical values found for the GaZNet morphoto-$z$ equivalent values in Table \ref{tab:masses_stats} ($\Delta p=-0.033$, NMAD$=0.093$ and Out. frac.$=3.8\%$). This quantifies the advantage of having accurate photo-$z$ in the stellar population analysis.
\end{appendix}

\end{document}